\documentclass[longauth]{aa}  
\usepackage{graphicx}
\usepackage{txfonts}
\usepackage[utf8]{inputenc}
\usepackage{array}
\usepackage{float}
\usepackage{multirow}
\usepackage{natbib}
\usepackage{epstopdf}
\usepackage{amsmath}
\usepackage[colorlinks=true,linkcolor=blue,citecolor=blue]{hyperref}

\newcommand{\mic}{$\mu$m\xspace}

 \hypersetup{draft}
\usepackage{color}

\begin{document}

   \title{SPHERE dynamical and spectroscopic characterization of HD142527B \thanks{Based on observations collected at the European Organisation for astronomical research in the southern emisphere under ESO programmes 095.C--0298, 096.C--0241, 097.C--0865 and 189.C--0209.}}

  % \subtitle{Something}

   \author{R. Claudi
          \inst{\ref{ist:oapd}}, %\fnmsep\thanks{Based on observations collected at Paranal Observatory, ESO (Chile), Program ID ... (SPERE)}
          %\and
         A.--L. Maire\inst{\ref{ist:mpiah}}, D. Mesa\inst{\ref{ist:oapd}, \ref{ist:inct} }, A. Cheetham\inst{\ref{ist:ogug}}, C. Fontanive \inst{\ref{ist:supa}}, R. Gratton \inst{\ref{ist:oapd}}, A. Zurlo\inst{\ref{ist:udpc}, \ref{ist:escuela}, \ref{ist:lam}}, H. Avenhaus\inst{\ref{ist:eth}, \ref{ist:dauc}, \ref{ist:dadisk}}, T. Bhowmik\inst{\ref{ist:lesiap}}, B. Biller\inst{\ref{ist:mpiah},\ref{ist:supa}}, A. Boccaletti\inst{\ref{ist:lesiap}}, M. Bonavita \inst{\ref{ist:oapd},\ref{ist:supa}}, M. Bonnefoy\inst{\ref{ist:ipag}}, E. Cascone\inst{\ref{ist:oacm}}, G. Chauvin\inst{\ref{ist:ipag},\ref{ist:mifca}}, A. Delboulb\'e\inst{\ref{ist:ipag}}, S. Desidera  \inst{\ref{ist:oapd}},  V. D'Orazi\inst{\ref{ist:oapd}}, P. Feautrier\inst{\ref{ist:ipag}}, M. Feldt\inst{\ref{ist:mpiah}}, F. Flammini Dotti\inst{\ref{ist:oapd}, \ref{ist:xian}}, J.H. Girard\inst{\ref{ist:ipag},\ref{ist:STSI}}, E. Giro \inst{\ref{ist:oapd}}, M. Janson\inst{\ref{ist:mpiah},\ref{ist:dasu}}, J. Hagelberg\inst{\ref{ist:ipag}}, M. Keppler\inst{\ref{ist:mpiah}}, T. Kopytova\inst{\ref{ist:mpiah},\ref{ist:soua}}, S. Lacour\inst{\ref{ist:lesiap}}, A.--M. Lagrange\inst{\ref{ist:ipag}}, M. Langlois\inst{\ref{ist:lam},\ref{ist:ensl}},  J. Lannier\inst{\ref{ist:ipag}},  H. Le Coroller\inst{\ref{ist:lam}}, F. Menard\inst{\ref{ist:ipag}}, S. Messina\inst{\ref{ist:oact}}, M. Meyer\inst{\ref{ist:eth}, \ref{ist:daum}}, M. Millward\inst{\ref{ist:ycotas}}, J. Olofsson\inst{\ref{ist:mpiah}, \ref{ist:IFA}, \ref{ist:NMFP}}, A. Pavlov\inst{\ref{ist:mpiah}}, S. Peretti\inst{\ref{ist:ogug}}, C. Perrot\inst{\ref{ist:lesiap}}, C. Pinte\inst{\ref{ist:ipag}, \ref{ist:moca}}, J. Pragt\inst{\ref{ist:nova}}, J. Ramos\inst{\ref{ist:mpiah}}, S. Rochat\inst{\ref{ist:ipag}}, L. Rodet\inst{\ref{ist:ipag}}, R. Roelfsema\inst{\ref{ist:nova}}, D. Rouan\inst{\ref{ist:lesiap}}, G. Salter\inst{\ref{ist:lam}}, T. Schmidt\inst{\ref{ist:lesiap}}, E. Sissa\inst{\ref{ist:oapd}}, P. Thebault\inst{\ref{ist:lesiap}}, S. Udry\inst{\ref{ist:ogug}}, A. Vigan\inst{\ref{ist:lam}}.
          }
   \institute{INAF -- Osservatorio Astronomico di Padova
              \email{riccardo.claudi@oapd.inaf.it}\label{ist:oapd}
\and Max Planck Institute for Astronomy, K\"onigstuhl 17, D--69117 Heidelberg, Germany \label{ist:mpiah}%2
\and INCT, Universidad De Atacama, calle Copayapu 485, Copiap\'{o}, Atacama, Chile \label{ist:inct}              
 \and Geneva Observatory, University of Geneva, Chemin des Maillettes 51, 1290 Versoix, Switzerland  \label{ist:ogug}%3
 \and SUPA, Institute for Astronomy, The University of Edinburgh, Royal Observatory, Blackford Hill, Edinburgh, EH9 3HJ, UK \label{ist:supa}%4
 \and N\'ucleo de Astronomia, Facultad de Ingenier\'ia, Universidad Diego Portales, Av. Ejercito, Santiago, Chile \label{ist:udpc}%5
\and Escuela de Ingenier\'ia Industrial, Facultad de Ingenier\'ia y Ciencias, Universidad Diego Portales, Av. Ejercito 441, Santiago, Chile \label{ist:escuela}%6
 \and Aix Marseille Universit\'e, CNRS, LAM (Laboratoire d'Astrophysique de Marseille) UMR 7326, 13388 Marseille, France  \label{ist:lam}%7
 \and ETH Zurich, Institute for Astronomy, Wolfgang-Pauli-Str. 27, CH-8093, Zurich, Switzerland \label{ist:eth}
 \and Departamento de Astronom\'ia, Universidad de Chile, Casilla 36-D, Santiago, Chile \label{ist:dauc}
 \and Millennium Nucleus ''Protoplanetary Disk'', Departamento de Astronom\'ia, Universidad de Chile, Casilla 36-D, Santiago, Chile \label{ist:dadisk}
 \and LESIA, Observatoire de Paris, PSL Research University, CNRS, Sorbonne Universit\'es, UPMC Univ. Paris 06, Univ. Paris Diderot, Sorbonne Paris Cit\'e, 5 place Jules Janssen, 92195 Meudon, France \label{ist:lesiap}%7
 \and Univ. Grenoble Alpes, CNRS, IPAG, F-38000, Grenoble, France \label{ist:ipag}%8
 \and INAF - Osservatorio Astronomico di Capodimonte, Salita Moiariello 16, 80131 Napoli, Italy \label{ist:oacm}
 \and Unidad Mixta Internacional Franco--Chilena de Astronomia, CNRS/INSU UMI 3386 and Departemento de Astronom\'ia, Universidad de Chile, Casilla 36--D, Santiago, Chile  \label{ist:mifca}%9
 \and Space Telescope Science Institute, 3700 San Martin Drive, Baltimore, MD 21218, USA \label{ist:STSI}
 \and European Southern Observatory (ESO), Alonso de C\'ordova 3107, Vitacura, Casilla 19001, Santiago, Chile \label{ist:esoc}   %10
 \and Department of Astronomy, Stockholm University, AlbaNova University Center, 106 91 Stockholm, Sweden\label{ist:dasu}   %11
\and Steward Observatory, The University of Arizona, Tucson, AZ 85721 \label{ist:soua}  %13
 \and CRAL, UMR 5574, CNRS, Universit de Lyon, Ecole Normale Superior de Lyon, 46 Alle d'Italie, F--69364 Lyon Cedex 07, France  \label{ist:ensl}%15
 \and INAF-Catania Astrophysical Observatory, via S. Sofia, 78, 95123, Catania, Italy \label{ist:oact}
 \and York Creek Observatory, Georgetown, 7253, Tasmania, Australia \label{ist:ycotas}
 \and Department of Astronomy, University of Michigan, 1085 S. University, Ann Arbor, MI 48109 \label{ist:daum}
 \and Instituto de F\'isica y Astronom\'ia, Facultad de Ciencias, Universidad de Valpara\'iso, Av. Gran Breta\~na 1111, Playa Ancha, Valpara\'iso, Chile \label{ist:IFA}
 \and N\'ucleo Milenio Formaci\'on Planetaria - NPF, Universidad de Valpara\'iso, Av. Gran Breta\~na 1111, Valpara\'iso, Chile \label{ist:NMFP}
 \and Monash Centre for Astrophysics (MoCA) and School of Physivs and Astronomy Monash University, Clayton Vic 3800, Australia \label{ist:moca}
 \and Xi'an Jiaotong--Liverpool University, department of Mathematical Sciences, 111 Ren?ai road, Suzhou Dushu Lake Higher Education town,Jiangsu Province, China, cp 215123\label{ist:xian}
 \and NOVA Optical Infrared Instrumentation Group, Oude Hoogeveensedijk 4, 7991 PD Dwingeloo, The Netherlands  \label{ist:nova}
             }

   \date{Received ....; accepted .....}

% \abstract{}{}{}{}{} 
% 5 {} token are mandatory
 
  \abstract
   {}
   {HD142527 is one of the most frequently studied Herbig Ae/Be stars with a transitional disk that hosts a large cavity that is up to about 100 au in radius.
  %probably the largest observed gap in any gas rich protoplanetary disk. 
  For this reason, it has been included in the guaranteed time observation (GTO) SpHere INfrared survey for Exoplanets (SHINE) as part of the Spectro-Polarimetric High-contrast Exoplanet REsearch (SPHERE) at the Very Large Telescope (VLT) in order to search for low-mass companions that might explain the presence of the gap. SHINE is a large survey within about 600 young nearby stars are observed with SPHERE with the aim to constrain the occurrence and orbital properties of the giant planet population at large ($>5$\ au) orbital separation around young stars.}
   {We used the IRDIFS observing mode of SPHERE (IRDIS short for infrared dual imaging and spectrograph plus IFS or integral field spectrograph) without any coronagraph in order to search for and characterize companions as close as 30 mas of the star. Furthermore, we present the first observations that ever used the sparse aperture mask (SAM) for SPHERE both in IRDIFS and IRDIFS\_EXT modes. All the data were reduced using the dedicated SPHERE pipeline and dedicated algorithms that make use of the principal component analysis (PCA) and reference differential imaging (RDI) techniques.}
   {We detect the accreting low--mass companion HD142527B at a separation of 73 mas (11.4 au) from the star. No other companions with mass greater than 10\ M$_J$ are visible in the field of view of IFS ($\sim 100$ au centered on the star) or in the IRDIS field of view ($\sim 400$ au centered on the star). Measurements from IFS, SAM IFS, and IRDIS suggest an M6 spectral type for HD142527B, with an uncertainty of one spectral subtype, compatible with an object of M$=0.11 \pm 0.06$\ M$_\odot$ and R$=0.15 \pm 0.07$\ R$_\odot$. The determination of the mass remains a challenge using contemporary evolutionary models, as they do not account for the energy input due to accretion from infalling material.  We consider that the spectral type of the secondary may also be earlier than the type we derived from IFS spectra. From dynamical considerations, we further constrain the mass to $0.26^{+0.16}_{-0.14}$\ M$_\odot$ , which is consistent with both our spectroscopic analysis and the values reported in the literature. Following previous methods,  the lower and upper dynamical mass values correspond to a spectral type between M2.5 and M5.5 for the companion. By fitting the astrometric points, we find the following orbital parameters: a period of P$=35-137$\ yr; an inclination of $i=121 - 130^\circ$\; , a value of $\Omega=124-135^\circ$\ for the longitude of node, and an 68\%  confidence interval of $\sim 18 - 57$\ au for the separation at periapsis. Eccentricity and time at periapsis passage exhibit two groups of values: $\sim$0.2--0.45 and $\sim$0.45--0.7 for $e$, and $\sim$2015--2020 and $\sim$2020--2022 for $T_0$.  While these orbital parameters might at first suggest that HD142527B is not the companion responsible for the outer disk truncation, a previous hydrodynamical analysis of this system showed that they are compatible with a companion that is able to produce the large cavity and other observed features.}
  {}


\titlerunning{SPHERE dynamical and spectroscopic characterization of HD142527B}
\authorrunning{Claudi et SPHERE GTO Cons.}
   \maketitle
%
%________________________________________________________________

\section{Introduction}
\label{sec:intro}   
%%%%%%%%%%%%
%V8.1 2018-12-17%%
%%%%%%%%%%%%

Planet formation from disks \citep{williamsandcieza2011, 2015IJAsB..14..201M} imprints characteristic structures on the disks \citep[e.g.,][]{2012ApJ...748L..22M}. One of the most striking structures that can be produced in this manner are the wide gaps carved by massive or multiple forming planets in protoplanetary disks \citep{2011ApJ...738..131D,2011ApJ...729...47Z}. Disks with large gaps, or cavities, are often referred to as ``transitional disks'', and they are observed around Herbig stars or their less massive counterparts, TTauri stars \citep{strometal1989}. 
A striking example of a transitional disk is the young Herbig Ae star (F6III, age$\sim 5$\ Myr) HD142527 \citep{Fukagawa2006} with its large gap. HD142527 shows a cavity as large as $\sim 100$\ au in radius. For this reason, it is one of the most freequently studied objects of this type (see Section\ \ref{sec:hd142527}).
Transitional disks are believed to be in the evolutionary stage between optically thick gas-rich disks and older disks where most of the gas has been dissipated (see, e.g., \citet{2014prpl.conf..497E} and references therein).
In a handful of cases the connection between the presence of a gap and the existence of (candidate) low-mass companions has been established. Some companion candidates were discovered in the LkCa 15 system \citep{2012ApJ...745....5K, 2015Natur.527..342S},  MWC\ 758 \citep{reggianietal2017}, HD169142 \citep{2014ApJ...792L..23R, 2014ApJ...792L..22B, osorioetal2014, 2017A&A...600A..72F, ligietal2018}, and HD100546 \citep{2013A&A...549A.112M, 2013ApJ...766L...1Q, 2015ApJ...807...64Q, 2013ApJ...767..159B, 2014ApJ...791..136B, 2015ApJ...807...64Q}. 
% \citep{2013A&A...549A.112M, 2013ApJ...766L...1Q, 2015ApJ...807...64Q, 2013ApJ...767..159B, 2014ApJ...791..136B, 2015ApJ...807...64Q, 2017AJ....153..264F, rameauetal2017}. 
HD142527 also has a low-mass companion that resembles an M star \citep{lacouretal2016}.
The recent advent of dedicated high-contrast imagers such as Spectro-Polarimetric High-contrast Exoplanet REsearch \citep[SPHERE,][]{beuzitetal2008}, Gemini Planet Imager \citep[GPI,][]{2014SPIE.9148E..0JM}, and Subaru Coronagraphic Extreme Adaptive Optics \citep[SCExAO,][]{2016SPIE.9909E..0WJ} offers the possibility to probe regions of transitional disk systems that lie closer to the star than was possible with previous imaging instruments and to constrain the presence of any companion that could be responsible for opening up an observed gap or cavity.

In this paper we present new deep images of the central regions of the HD142527 system obtained with SPHERE as part of the SPHERE consortium guaranteed time observations (GTO). These data include both non-coronagraphic direct images and sparse aperture masking (SAM) data acquired with the near-infrared channels integral field spectrograph (IFS) and infra-red dual imaging and spectrograph (IRDIS). The outline of the paper is as follows: in Sect. \ref{sec:hd142527} we summarize the main characteristics of the HD142527 system. In Sect. \ref{sec:obs} we describe the SPHERE near-infrared observations; in Sect. \ref{sec:dr} we describe the reduction methods we applied, and the results are discussed in Sect. \ref{sec:res}. The mass estimate of the companion is discussed in Sect.\ \ref{sec:massB}, and  the orbital properties of HD142527B are derived in Sect. \ref{sec:orbit}. In Sect. \ref{sec:disc} we outline the conclusions.

\begin{table*}
\caption{HD142527 Observations. }           
\label{table:1}     
%\centering                         
\begin{tabular}{c c c c c c c c c c} 
\hline\hline                
UT Date & Instr. Mode & Instr.& Filter & DIT$\times$NDIT$^{\bf a}$& N$_{exp}^{\bf a}$& Field Rot. & Seeing & Strehl              & True North\\    % table heading 
              &                    &         &          &                             &                  &                  &             &                         & Correction\\
              &                    &         &          & (s)                        &                  & ($^\circ$)  &  ('')        & @$1.6 \mu$m &  ($^\circ$) \\
\hline                        % inserts single horizontal line
\multirow{4}{*}{2015-05-13} & \multirow{4}{*}{IRDIFS\_EXT}  &  IFS    & YJH/ND1         & 4.0 $\times$56   & 8  &\multirow{4}{*}{40.10}    &    \multirow{4}{*}{0.3} &  \multirow{4}{*}{0.8}     &  \multirow{4}{*}{$-1.86 \pm 0.15$}  \\
 &   &  IRDIS& K1--K2/ND1  & 0.84 $\times$88            & 16  &      &      &    &   \\
 &   &  IFS    & YJH/ND2         & 8.0 $\times$8          & 8  &      &      &    &   \\
 &   &  IRDIS& K1--K2/ND2  & 8.0 $\times$8   & 8 &      &      &    &   \\
 \hline
 \multirow{2}{*}{2015-07-03}& \multirow{2}{*}{SAM}& IFS &YJ   &4 $\times$ 8    &3   &   \multirow{2}{*}{25.00}&\multirow{2}{*}{0.5}& \multirow{2}{*}{0.7}& \multirow{2}{*}{$-1.768 \pm 0.055$}\\ 
                                             &                                 & IRDIS&H2--H3    &0.84 $\times$32    &12   &&                          &                              &       \\
\hline
\multirow{4}{*}{2016-03-26} &  \multirow{4}{*}{IRDIFS\_EXT}  &  IFS& YJH/ND1  &4.0 $\times$56& 16     &\multirow{4}{*}{73.84}& \multirow{4}{*}{0.7} & \multirow{4}{*}{0.7} &\multirow{4}{*}{$-1.756 \pm 0.061$}     \\
                                            &                                                    & IRDIS  & K1--K2/ND1 &0.84 $\times$88 & 32 &      &       &         &     \\
                                            &                                                    & IFS  & YJH/ND2 &8.0 $\times$8 & 12 &         &       &               &     \\
                                            &                                                    & IRDIS  & K1--K2/ND2 &2.0$\times$19 & 12 &         &       &               &     \\
\hline                                             
\multirow{2}{*}{2016-06-13} & \multirow{2}{*}{IRDIFS\_EXT}  &IFS& YJH/ND1  & 4.0$\times$60&16 &\multirow{2}{*}{64.48}&\multirow{2}{*}{0.8}&\multirow{2}{*}{0.6}  & \multirow{2}{*}{$-1.664 \pm 0.048$}     \\
                                            &                                                    & IRDIS  & K1--K2/ND1 &0.84 $\times$49 & 64 &      &       &         &     \\
\hline
\multirow{3}{*}{2017-05-16} & \multirow{3}{*}{SAM}  &IFS& YJH  & 4.0$\times$16&24 &\multirow{3}{*}{73.00}&\multirow{3}{*}{0.6}&\multirow{3}{*}{0.7}  & \multirow{3}{*}{$-1.800 \pm 0.052$}     \\
                                            &                                                    & IRDIS  & K1--K2 &0.84 $\times$32 & 4 &         &       &               &     \\
                                            &                                                    & IRDIS  & K1--K2 &0.84 $\times$52 & 20 &         &       &               &     \\
\hline                                                                                         
\multirow{2}{*}{2018-04-14} & \multirow{2}{*}{SAM}  &IFS& YJH  & 2.0$\times$18&16 &\multirow{2}{*}{52.00}&\multirow{2}{*}{0.6}&\multirow{2}{*}{0.9}  & \multirow{2}{*}{$-1.75 \pm 0.10$}     \\
                                            &                                                    & IRDIS  & K1--K2 &0.84 $\times$36 & 16 &         &       &               &     \\
\hline                                                                                         
\hline                                   %inserts single line
\end{tabular}
\\[5pt] $^a$  NDIT refers to the number of integrations per datacube, DIT to the integration time, N$_{exp}$ to the number of datacubes 
\end{table*}

\section{HD142527}
\label{sec:hd142527}   
HD142527 is a young  $5 \pm 1.5$ Myr \citep{2014ApJ...790...21M} intermediate-mass star at a distance of $140 \pm 20$\ pc  from the Sun \citep{2004A&A...426..151A,2014ApJ...790...21M}.  Its spectral type of F6 IIIe \citep{1978mcts.book.....H,1976ApJS...30..491H,1996A&A...315L.245W} corresponds to a mass of $2.0 \pm 0.3$ M$_\odot$, \citep{Fukagawa2006,Verhoeff2011}. 
%The star's proper motion of $\mu_\alpha = -17.2$ mas/yr, $\mu_\delta = -18.0$ mas/yr \citep{2007A&A...474..653V} is similar to that of stars associated with the Lupus clouds \citep[$\mu_\alpha = -16.0$ mas/yr, $\mu_\delta = -21.7$ mas/yr;][]{2013A&A...558A..77G}, thus the star is likely to be co--distant with either the Lupus clouds \citep{2013A&A...558A..77G} or the Upper Cen-Lup subgroup of Sco-Cen \citep[$\sim142$ pc][]{1999AJ....117..354D} which surrounds the Lupus clouds. 
%Recently, \citet{GaiaCollaboration2016} refined the distance to $156^{+7}_{-6}$\ pc and the proper motion to $\mu_\alpha=-11.76 \pm 0.08$\ mas/yr, $\mu_\delta=-24.45 \pm 0.05$\ mas/yr.
We adopt the distance recently refined by \citet{GaiaCollaboration2016} to $156^{+7}_{-6}$\ pc and the proper motion of $\mu_\alpha=-11.76 \pm 0.08$\ mas/yr, $\mu_\delta=-24.45 \pm 0.05$\ mas/yr.

HD142527 is one of the most frequently studied Herbig Ae/Be stars with a transitional disk because its disk is seen almost face-on and its protoplanetary cavity, extending to a radius of about 100 au, can be investigated because of its record size \citep{Fukagawa2006}.
%has probably the largest imaged gap in any protoplanetary disk. 
The gap in this system extends to between 30 and 130 au \citep{Verhoeff2011}, and the outer disk extends to about $600$\,au. The system is seen at a low inclination: $i = 28 \pm  3^\circ$ \citep{perezetal2015}. The HD142527 disk has long been posited as a possible site of planet formation because of the extremely high fraction of crystalline silicates, which have possibly formed by a massive companion that induced spiral density waves in the disk material \citep{2004Natur.432..479V, avenhausetal2017}. \citet{Fukagawa2006} imaged in scattered light the outer edge of the gap as well as a spiral feature in the outer disk.  These features have recently been confirmed in the visible polarized light by \citet{avenhausetal2017}.
The protoplanetary disk of HD142527 shows high near-infrared excess \citep{1999A&A...345..181M,Fukagawa2006} that indicates that optically thick material lies close to the star; this might be a remnant of the original inner disk. The central star still accretes at a rate of $9.5 \times 10^{-8}$ M$_\odot$/yr \citep{garcialopezetal2006,Casassus2012,Avenhaus2014,2012ApJ...753L..38B,2013ApJ...769...21S}, suggesting that there is still enough material close to the star for it to be funneled onto the star. The outer radius of this inner disk is likely within $10$\,au \citep{Verhoeff2011,2013PASJ...65L..14F, closeetal2014}.  \citet{Fukagawa2006} also found an offset of 20\ au between the location of the star and the center of the disk, which may be caused by an unseen (at the epoch) eccentric binary companion. The gap appears to be completely depleted of both large and small dust grains, as evidenced from scattered light images as well as (sub-)\,millimeter observations. Nonetheless, CO gas has been detected within the gap \citep{2013Natur.493..191C}.

Using SAM with NACO \citep[short for NAOS-CONICA or Nasmyth adaptive optics system near infrared imager and spectrograph,][]{2010SPIE.7735E..1OT} at the Very Large Telescope (VLT), \citet{2012ApJ...753L..38B} detected an asymmetry in the brightness distribution around the central star with a barycenter emission located at a projected separation of $88 \pm 5$ mas ($12.8 \pm 1.5$ au at 145 pc) and flux ratios in H,K, and L' of $0.016 \pm 0.007$, $0.012 \pm 0.008$, and $0.0086 \pm 0.0011$, respectively ($3\sigma$ errors), relative to the primary star and disk. They interpreted this asymmetry as a low-mass stellar companion ($\sim 0.2$ M$_\odot$) orbiting at $\approx12 $ au from the star, well inside the gap. On the basis of their observations with near infrared coronagraphic imager (NICI) at Gemini south, \citet{2013Natur.493..191C} disputed the presence of a companion, but  \citet{closeetal2014} confirmed the companion's existence through direct-imaging observations in the R and in H$_\alpha$ bands. The latter implies mass accretion, and the authors quantified it as $\sim 5.9 \times 10^{-10}$\ M$_\odot$ yr$^{-1}$. Recently, \citet{lacouretal2016} observed HD142527 from $R-$ to $M$--band wavelengths with the NACO and Gemini Planet Imager \citep{2014SPIE.9147E..7BG} instruments using the SAM technique. They constrained the companion mass and radius with evolutionary models as $0.13 \pm 0.03$ M$_\odot$ and $0.90 \pm 0.15$ R$_\odot$ , respectively, and derived a  younger age ($1.0^{+1.0}_{-0.75}$\ Myr) than for HD142527A. The characteristics of this system, and in particular the existence of the wide gap beyond the low-mass companion, make the HD142527 system a prime target for the search for circumbinary planets \citep{bonavitaetal2016}. 

\section{Observations}
\label{sec:obs}
The SPHERE planet-finder instrument installed at the VLT \citep{beuzitetal2008} is a highly specialized instrument dedicated to high-contrast imaging at optical and near-infrared wavelengths. It is equipped with an extreme adaptive optics system called SAXO (Sphere Adaptive Optics for eXoplanet Observation) \citep{fuscoetal2014, 2014SPIE.9148E..0OP}, with a $41 \times 41$ actuator wavefront control, pupil stabilization, and differential tip-tilt control. It also employs stress-polished toric mirrors for beam transportation \citep{2012A&A...538A.139H}. The SPHERE instrument is equipped with several coronagraphic devices for stellar diffraction suppression, including apodized Lyot coronagraphs \citep{2005ApJ...618L.161S} and achromatic four-quadrant phase masks \citep{2008SPIE.7015E..1BB}. The instrument has three science subsystems: the infrared dual-band imager and spectrograph \citep[IRDIS;][]{2008SPIE.7014E..3LD}, an integral field spectrograph \citep[IFS;][]{2008SPIE.7014E..3EC, 2016SPIE.9908E..3HC}, and a rapid-switching imaging polarimeter \citep[ZIMPOL;][]{2008SPIE.7014E..3FT}. \par
Our observations were part of the SHINE \citep[SpHere INfrared survey for Exoplanets,][]{chauvinetal2017} survey and were performed with SPHERE in the IRDIFS\_EXT mode (direct imaging). In this mode, IRDIS observes in dual-band imaging
\citep[DBI;][]{2010MNRAS.407...71V} with the K12 filter pair (wavelength K1=2.110 \mic; K2=2.251 \mic), while IFS obtains low-resolution (R=30) spectra between 0.95 and 1.65 \mic. The target has been observed with SAM both in IRDIFS and IRDIFS\_EXT mode.
In contrast to the IRDIFS\_EXT mode, in the IFS mode IRDIS observes in dual-band imaging with H23 filter pair (wavelength H2=1.589 \mic; H3=1.667 \mic), while IFS performs low-resolution spectroscopy at R$\sim 50$ in the wavelength range 0.95 -- 1.35 \mic.

\begin{figure*}%[!htp]
\begin{center}
\centering
\includegraphics[width=16.0cm]{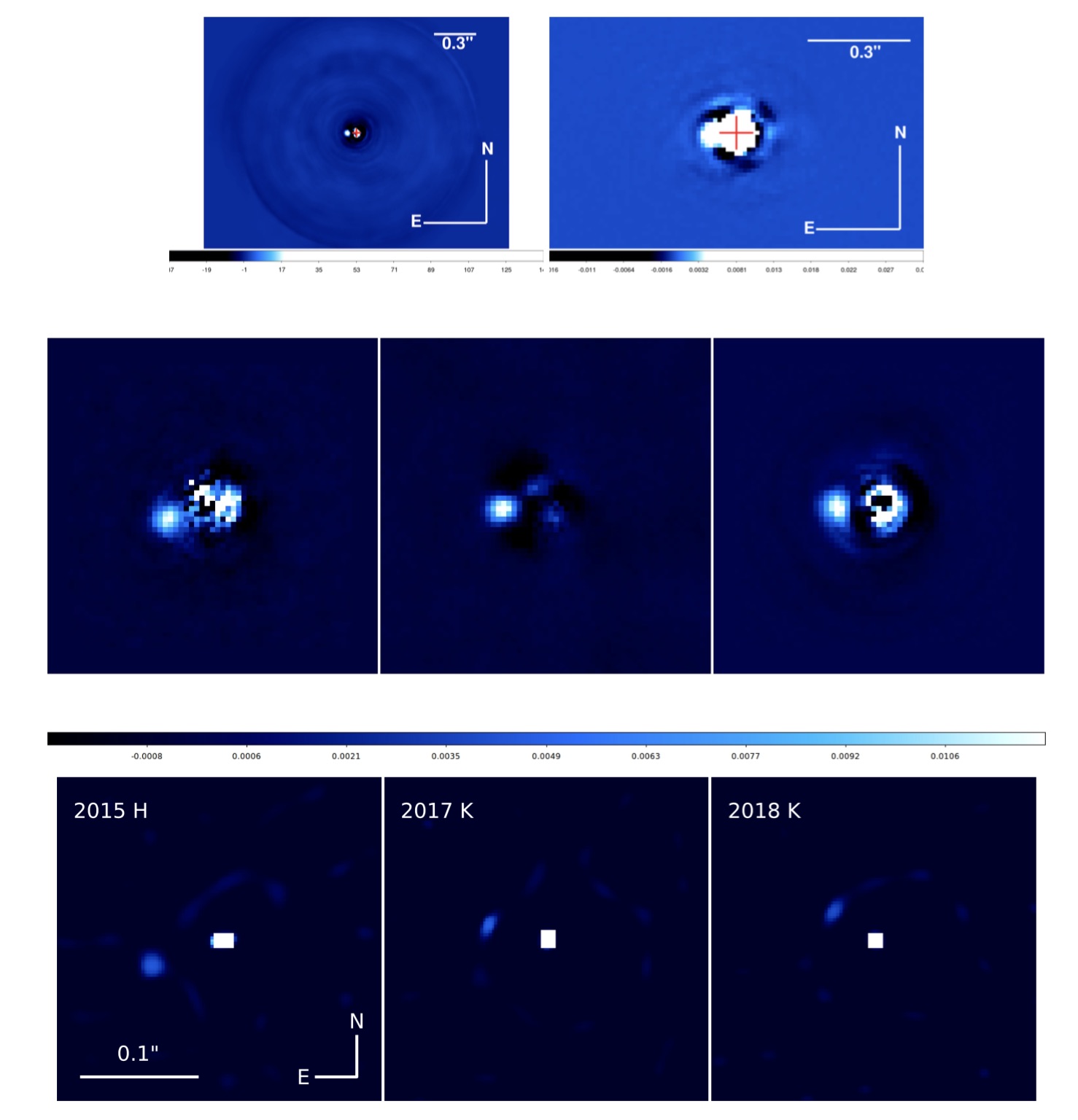}
\caption{Non-coronagraphic IRDIFS\_EXT and SAM images of the HD142527 system. {\bf Upper panel}: Images  acquired on June 13,$^{}$ 2016. {\bf Left}: PCA post-processing IFS image (averaged over all channels).  The red cross marks the position of the central star. In this image the companion of HD142527 is clearly visible. {\bf Right}: Central part of the PCA post-processing IRDIS image of HD142527.  {\bf Central panel}: Composition of the three RDI post-processing IFS images (from left to right: May 2015, March 2016, and June 2016). The orbital motion of B is clearly detected. {\bf Bottom panel}: Reconstructed images produced from the SAM data using the MiRA algorithm. The images show that we observe significant orbital motion for the companion HD 142527B between the SAM epochs, and that the point-source model used to calculate its position is a good approximation to the observed structure}
\label{f:imgifs}
\end{center} 
\end{figure*}

\subsection{Direct imaging}
We observed HD142527 during five nights (2015--03--13, 2016--03--26, 2016--06--13, 2017--05--16, and 2018--04--14) as part of the SPHERE GTO program (IRDIFS and SAM) plus one in technical time dedicated to the commissioning of the SAM observing mode (2015--07--03). The instrumental setups used during these observations are all described in Table\ \ref{table:1}.
The main goal of these observations was to detect and characterize the stellar companion around HD142527 and to place strong constraints on the presence of other possible companions. Because the star -- companion separation is expected to be smaller than ($88 \pm 5$\  mas), we observed the star without a coronagraph,
and adjusted the exposure time in order to avoid detector saturation. In the first two observations with the modified IRDIFS\_EXT instrument mode (no coronagraph), the IRDIS images taken with the 1/10 neutral density filter (ND1) were slightly saturated.
For this reason we took additional unsaturated exposures using the 1/100 neutral density (ND2) filter to properly calibrate the flux from the star and ND1 images. No saturation was present in the IFS images, so all IFS images could be used as science images.  The observing conditions were worse for the IRDIS observations, which are not saturated on the third night, so that we were able to use only the ND1 filter to take our science images. 
In order to attenuate the residual speckle noise with the angular differential imaging \citep[ADI;][]{2006ApJ...641..556M} technique in the post-processing phase, we acquired our observations in pupil-stabilized mode.

\subsection{Sparse aperture masking} % ACC new text
HD142527 was also observed in the SAM mode of SPHERE \citep{2016SPIE.9907E..2TC} on the night of March 7, 2015, during a technical night to test this observing mode, and additionally, as part of SHINE, on the nights of May 16, 2017, and April 14, 2018. This was the first time in which the SAM mode was used with SPHERE. The object was observed in IRDIFS (July 3, 2015) and in IRDIFS\_EXT (May 16, 2017 and April 14, 2018) modes with a seven-hole pupil-aperture mask with the same layout as was used with NACO. The observing conditions during the 2015 observation were poor, with thick clouds, low-coherence times, and high values of seeing affecting the performance of the AO system. 

To calibrate the systematic effects that are present in SAM (and similar interferometric techniques), the calibrator stars HD 142695 and HD 142277 were also observed. The observations used 0.84\,s and 8\,s exposure times for IRDIS and IFS, respectively (see\ Table\ \ref{table:1}). 
%A total of 96 IRDIS images and 24 IFS images were obtained for the target, with 32 IRDIS images and 8 IFS images taken for each of the calibrators.

\section{Data reduction and analysis}
\label{sec:dr}
The data gathered in the six observation nights have been reduced following the necessary data reduction recipes for the individual instruments in both the classical (IRDIFS) mode and IFS and IRDIS in SAM mode. In this paragraph the data reduction is discussed for each instrument.
\subsection{IFS data}
For the IFS data we began with the reduction of calibration data (dark, detector flat, spectral position frames, wavelength calibration, and instrument flat) using the data reduction and handling (DRH) software \citep{2008SPIE.7019E..39P}. For a more detailed description of each of these steps, we refer to \citet{mesaetal2015}. 
The raw science data frames were then averaged so that the rotation between any two frames was on the order of 0.3 degrees.  This left us with 128 different frames for the first night and 248 for the
second night, when a larger dataset was obtained. For the third night, we had 96 frames after binning.
For each of these frames, we then corrected the bad pixels and the effects of the cross-talk between different lenslets of the IFU \citep{2009ApJ...695.1042A} using dedicated IDL procedures described in more detail in \citet{mesaetal2015}. 
On each frame, we then ran the DRH science recipe, which uses the calibration files and produces a wavelength-calibrated datacube composed of 39 monochromatic images. 
Each of these images was then corrected for the different exposure times and neutral density filters used. We then determined the position of the star for each frame using the CNTRD\footnote{http://www.harrisgeospatial.com/docs/cntrd.html} IDL procedure and recentered each of them to the nominal position of the image center. 
Finally, to retrieve the companion image, we performed a principal component analysis (PCA) procedure as described in \citet{mesaetal2015}, exploiting both the angular and spectral information. As we observed the object in a non-coronagraphic mode and to reduce concern for the self-cancellation that is present in the PCA method, we also performed a point spread function (PSF) subtraction (reference differential imaging, RDI) in a region very close to the star. We devised an automatic search in the SPHERE Data Center \citep[SDC,][]{delormeetal2017} for reference sequences obtained using the same observing mode (non-coronagraphic observation in IRDIFS\_EXT mode). We found that the best match was an observation of HD100546, which provides the highest correlation with the observation of HD142527, taken in February 2017,  although this star also hosts a bright disk. This sequence includes 533 individual 3D (x, y, lambda) datacubes. We are aware that HD100546 has a bright disk that is well visible in scattered light (see, e.g., \citet{sissaetal2018}). However, first, the model is obtained by an automatic procedure. Second, the inner disk of HD 100546 is not resolved in SPHERE images and the apparent separation of the inner edge of the intermediate ring is at about twice that of the companion of HD142527. Hence, the disk of HD 100546 has no impact on the present discussion. 
For each separate wavelength image, we first accurately recentered each image on the peak of the diffraction image. For each monochromatic image of the science datacube, our code then searched for the monochromatic image of the reference datacube that provides the highest cross-correlation within a circle with a radius of 12 pixels from the nominal center. This is selected as the best -matching monochromatic image. The flux of the best-matching monochromatic image is normalized to the  value of the science monochromatic image and then subtracted. The subtracted image is normalized to  the peak of the original image, thus we therefore obtain a contrast image. The subtracted monochromatic image is  derotated and a median of the derotated monochromatic images is then made over time. A high-pass filtering is then made by subtracting the current median over an area of $21 \times 21$ pixels centered on each pixel. The final images are then obtained by collapsing the datacubes along the spectral axis.

\subsection{IRDIS data}
Data reduction for the IRDIS observations was performed following the procedures described in \citet{zurloetal2014, 2016A&A...587A..57Z}. The IRDIS raw images were pre-reduced by performing background subtraction, bad-pixel correction, and flat fielding.  As these are non-coronagraphic images, no satellite spots or PSF reference images were taken. Satellite spots are fiducial spots symmetric with respect to the central star created by using a periodic grid mask downstream a coronagraph \citep{sivaramakrishnanandoppenheimer2006, maroisetal2006} or by applying a periodic modulation on an adaptive optics deformable mirror \citep{langloisetal2013}. We used one of the images in the sequence as a reference in order to apply the SDC data reduction procedure. The frames within the resulting datacube were then aligned with one another after finding for each of them the location of the star using the CNTRD IDL procedure and recentering each of them to the nominal position of the image center. For all epochs, after the preprocessing of each frame, the speckle pattern subtraction was performed using both the PCA \citep{soummeretal2012} and the TLOCI \citep{2014SPIE.9148E..0UM} algorithms, combined with the ADI technique. 

\subsection{Sparse aperture masking data analysis} % ACC new text
The SAM IFS data were converted into cleaned wavelength-extracted cubes using the SPHERE DRH in the same manner as the coronagraphic frames, but without any data binning.
The IRDIS and IFS wavelength cubes were then processed using the IDL-based aperture masking pipeline developed at the University of Sydney, with recent modifications allowing the simultaneous processing of multiwavelength data. A more thorough description of the pipeline can be found in \cite{2000PASP..112..555T}, \cite{2008ApJ...679..762K}, and the references therein, but a brief summary follows.

The data were background subtracted, flat fielded, and cleaned of bad pixels and cosmic rays. The cleaned cubes were then windowed with a super-Gaussian function of the form $e^{(-ar^4)}$ before closure phases were extracted from the Fourier transforms of the images. Calibration of the closure phases was performed on each wavelength individually by subtracting a weighted sum of the corresponding measurements taken on the calibrator stars.

The calibrated closure phases were fit with a binary analytical model with the following free parameters: the separation, position angle, and a contrast value for each wavelength channel. The IFS and IRDIS data were fit separately, resulting in four free parameters for IRDIS and 41 parameters for IFS. The best-fit parameters were estimated using {\it emcee} - a Python implementation of the affine-invariant Markov chain Monte Carlo (MCMC) ensemble sampler \citep{2013PASP..125..306F}. To account for uncalibrated systematics, a constant was added in quadrature to each closure phase uncertainty to ensure that the best-fitting model had a reduced $\chi^2$ of 1. These constants were estimated to be $0.45^{\circ}$ and $0.4^{\circ}$ for IRDIS and IFS, respectively, in comparison to initial median uncertainties of $0.3^{\circ}$\, and $0.4^{\circ}$. These high values suggest that the imperfect calibration dominates the uncertainties.

To estimate the detection limits from the observations, a Monte Carlo simulation was performed. We generated a set of 10,000 simulated datasets drawn from a Gaussian distribution, consistent with the measured uncertainties. For each point on a grid of separation and contrast, our detection limits were defined as the point at which at least 99.9\% of the datasets were fit better by a point-source model than the binary model. The 99.9\% criteria yields a set of 3.3$\sigma$ detection limits, which were approximately 7.5\,mag for separations between 50-250\,mas.

\section{Results}
\label{sec:res}

The final non-coronagraphic images for IFS, IRDIS, and SAM are shown in Figure~\ref{f:imgifs}. Images are from the last night of IRDIFS\_EXT observations; similar images were obtained on the other two nights when the system was observed with the same instrumental mode. The companion is clearly visible in the IFS images with a signal-to-noise ratio (S/N) 
%We calculated the signal from the object with an aperture photometry on an $0.5\lambda/D$ aperture. The noise was evaluated in two zones at the same separation from the central star of the object and with separations between 2.5 and $6 \lambda/D$ from the object itself.} 
of $\sim$30 \citep[the procedure used to evaluate the S/N is fully described in][]{zurloetal2014}, but it is only marginally resolved in the IRDIS frames because of the very small angular separation of the companion and the less favorable pixel scale of IRDIS. Therefore, we did not use the IRDIS non-coronagraphic data for the astrometric and photometric characterization of HD142527B. The central panel of the same figure shows a composition of three post-processing IFS images of the HD142527 system taken at different epochs (May 2015, March 2016, and June 2016). The companion is also clearly visible in the SAM data with an S/N greater than $60$ with IRDIS and greater than $40$ with IFS. The S/N in the SAM observations was calculated by comparison of the best-fit flux ratio with the detection limits.We used the MiRA image reconstruction algorithm \citep{2008SPIE.7013E..1IT} to produce the images in the bottom panel of Figure \ref{f:imgifs}. MiRA uses an inverse-problem approach to reconstruct an image from the limited information provided by the closure phases and power spectrum. The images confirm that the data are consistent with a binary companion and show its orbital motion between the IFS and SAM datasets.
% The images clearly show HD142527B and we can already see small displacements of the companion due to its orbital motion. 

\begin{figure}%[!htp]
\begin{center}
\centering
\includegraphics[width=9.0cm]{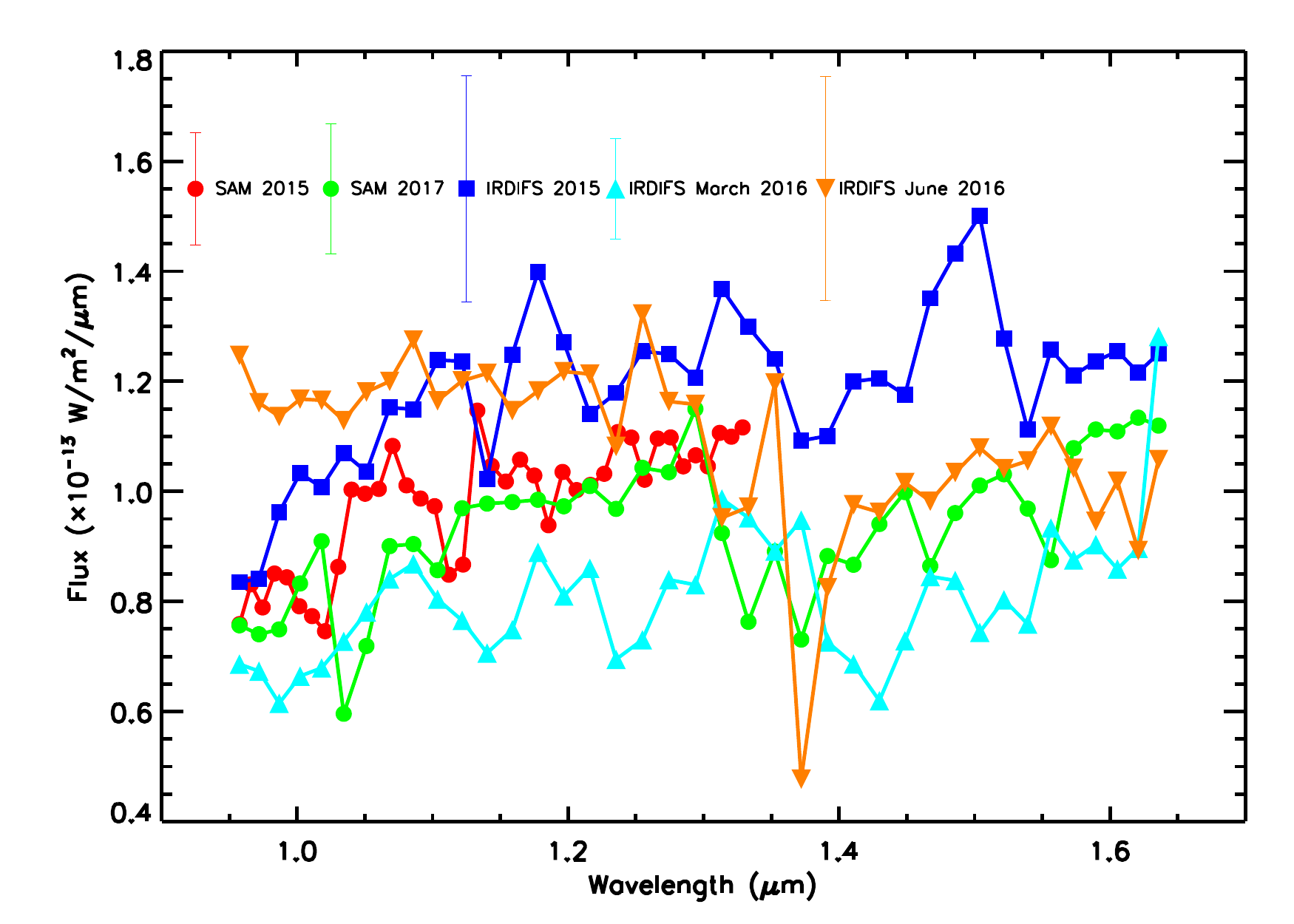}
\caption{Comparison among the extracted spectra for HD142527B obtained by both direct imaging and SAM modes of SPHERE in the IFS wavelength range ($0.95 - 1.65\ \mu$m). In the upper part of the plot is indicated the value of the average value of the error for each spectrum. The SAM IFS spectrum taken in 2015 is limited to the YJ band of  IFS ($0.95 - 1.35\ \mu$m). %The spectra at different epochs show slight variability in flux. 
For a clearer view of the plot all points are connected by continuous lines.}
\label{f:comp}
\end{center} 
\end{figure}

\subsection{HD142527B spectrum}
\label{sec:Spec}
Figure~\ref{f:imgifs} shows that the PCA and RDI post-processing techniques allow us to very clearly detect HD142527B in IFS data. However, at this small separation, self-subtraction will severely diminish the flux in the spectrum of the companion. 
To avoid self-subtraction due to PCA, we calculated photometry from each IFS spectral channel and the corresponding errors from images obtained with RDI, exploiting the method described in \cite{zurloetal2014}.
%with a simple stellar radial profile subtracted. In these images the companion is barely visible, but we can adopt for the companion the position calculated using the images reduced with the PCA algorithm and thus still extract accurate photometry. 
All the extracted spectra (for IFS and SAM) are shown together in Figure\ \ref{f:comp}. In order to facilitate comparison between the spectra taken at different epochs, we plot only points falling in the IFS wavelength range ($0.95-1.65\ \mu$m). The H$_2$ H$_3$ and K$_1$ K$_2$ points from SAM observations are not shown, but are listed in Table\ \ref{table:photo}.

\begin{table*}
\caption{Absolute magnitudes of HD142527\ B obtained from the IFS data for Y, J, and H spectral bands. SAM data also contain the H$_2$, H$_3$ and K$_1$ and K$_2$ bands of IRDIS . These magnitudes are not corrected for the several reddening contributions from the inner part of the HD142527 system  (see text).}             
\label{table:photo}    
\centering                         
\begin{tabular}{c c c c c c c c} 
\hline\hline                
Date & Y & J  & H & H$_2$  &H$_3$ & K$_1$ & K$_2$ \\    % table heading 
\hline                        % inserts single horizontal line
2015-05-13 & 5.19$\pm$0.22 & 4.91$\pm$0.16 & 4.01$\pm$0.17 & & \\
2015-07-03 & 5.38$\pm$0.12 & 5.09$\pm$0.09 &                          & 3.98$\pm$0.03  &3.93$\pm$0.03   \\
2016-03-26 & 5.59$\pm$0.11 & 5.33 $\pm$0.09 & 4.44$\pm$0.12 & &   \\
2016-06-13 & 5.07$\pm$0.22 & 4.99$\pm$0.18 & 4.22$\pm$0.17 & &   \\
2017-05-16 & 5.47$\pm$0.13 & 5.20$\pm$0.14 & 4.27$\pm$0.10 & &  &3.72$\pm$0.01 &  3.71$\pm$0.02 \\
2018-04-14 & &  &  & &  &3.69$\pm$0.03 &  3.74$\pm$0.04 \\
\hline  
\\                                 %inserts single line
\end{tabular}
\end{table*}

\begin{figure}%[!htp]
\begin{center}
\centering
\includegraphics[width=6.5cm]{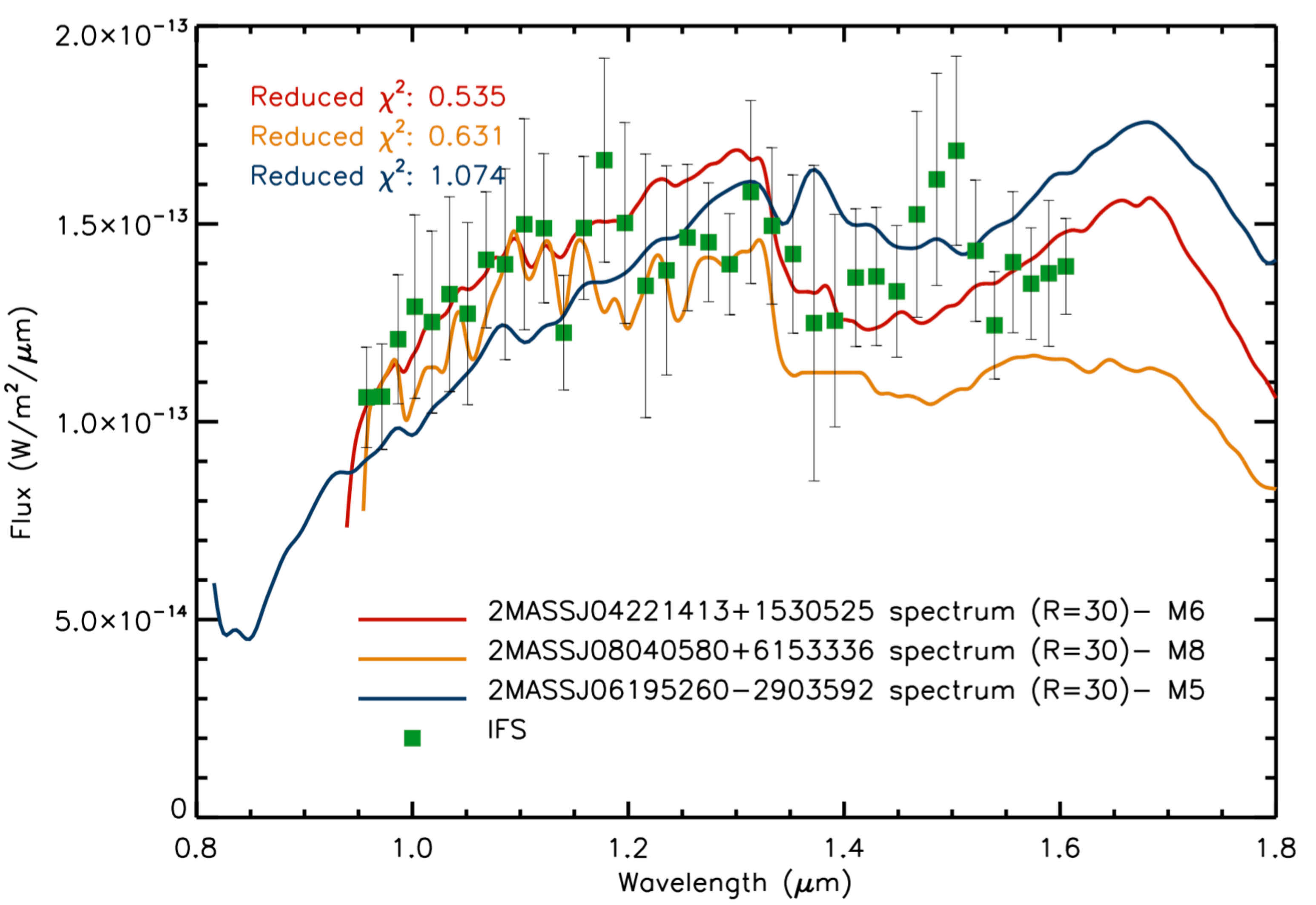}
\includegraphics[width=6.5cm]{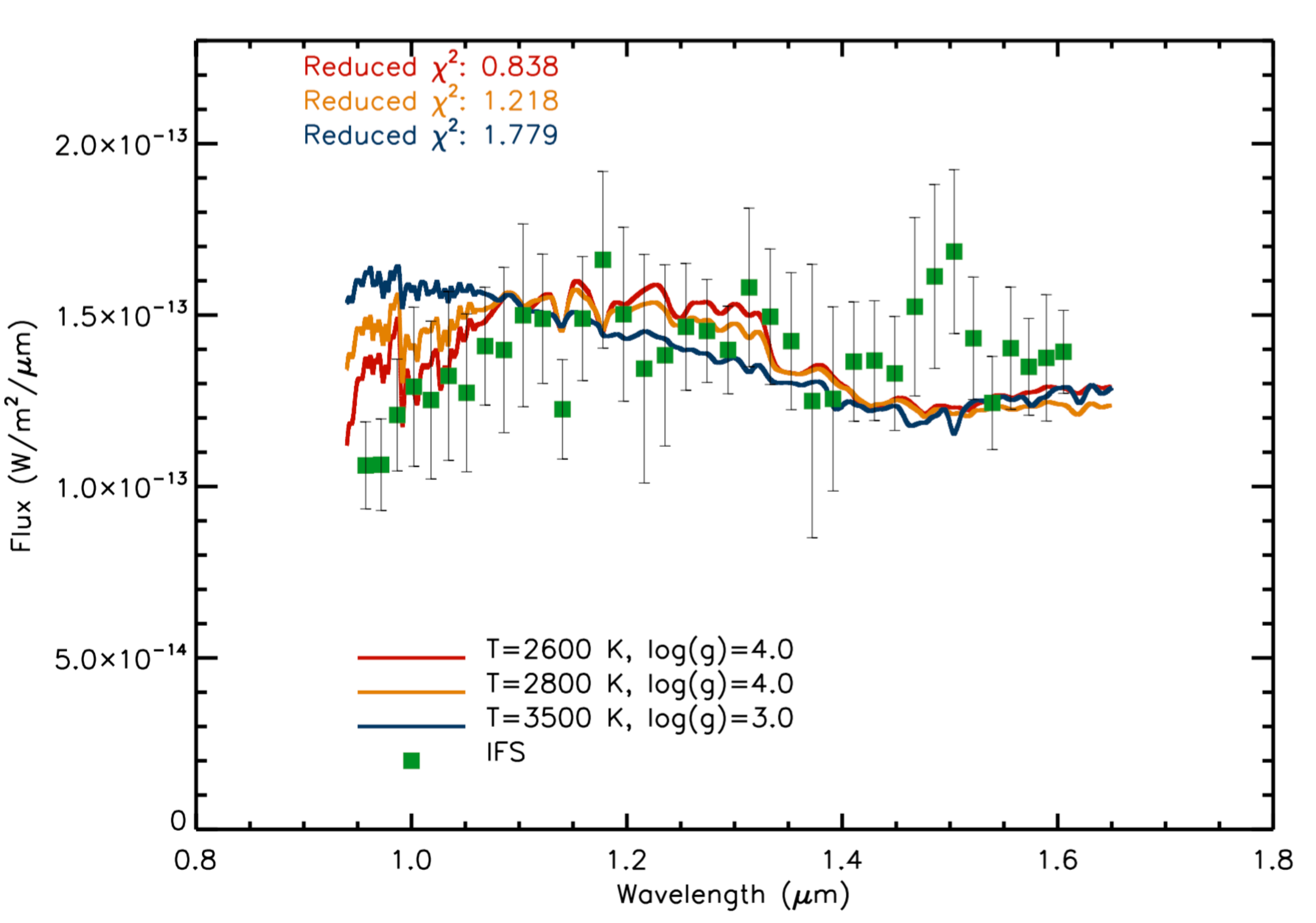}
\caption{{\bf Top}: Extracted spectrum for HD142527B (green squares) compared with the three best-fit spectra (red, orange, and blue solid lines)
from the \citet{2013ApJ...772...79A} obtained from the data taken on the night of May 13, 2015. {\bf Bottom}: Same as the left panel, but compared with the three best-fit models (red, orange, and blue solid lines) from the BT-Settl library \citet{allard2014}.}
\label{f:spfit1}
\end{center} 
\end{figure}

Each extracted spectrum of HD142527B  can be fit with spectra of young field dwarfs from \citet{2013ApJ...772...79A} to estimate its spectral type. This procedure was executed for all the epochs, and the results are shown in Table\ \ref{table:spec}.  The low reduced $\chi^2$ values ($<1$) that we obtain in some fits are probably due to the combination of two effects: {\it i}) all the errors could be overestimates, and {\it ii}) some of the 39 measurements of each spectrum could have a certain covariance degree between closer spectral channels. The fit is useful, however, to form an impression of  the spectral type of the secondary. Figure\ \ref{f:spfit1} shows the best-fitting spectra overlaid on the IFS spectrum from May 14, 2015. 
In order to constrain the physical characteristics of the companion, we compared the extracted spectra with a set of BT-Settl models \citep{allard2014}. The models were selected in a grid with the effective temperature and the surface gravity ranging in the following intervals: $1000 \leq$ T$_{eff} \leq 4000$\ K and $3.0 \leq \log(g) \leq 5.5$, with an incremental step of 100\ K for the former and 0.5 for the latter. In all cases the fit with models with T$_{eff}=2600\ -\ 2800$\ K  and $\log(g) \sim 4.0$ was good, also when higher temperatures and lower surface gravities were included. As an example of the obtained results, we show in the right panel of Figure \ \ref{f:spfit1} the models and in the left panel, we show the same, but fit with the best-fit spectra.

The fit results for a range of spectral types are presented in Table\ \ref{table:spec}. The best-fit spectra match well at all epochs (except for the spectrum from June 13, 2016), thus we determine a spectral type between M5 and M6 for HD142527B with an uncertainty of $\pm1$ subtype, in agreement with the model fitting. From \citet{pecautandmamajek2013} we obtained the value of the T$_{eff}$ corresponding to the best-fit spectral type for each individual observation (Col. 3 of Table\ \ref{table:spec}). 
The extracted spectra seem to be different from each other (see Figure\ \ref{f:comp}). We evaluated the Y--H color of each spectrum using the value of the magnitude discussed in the next section and reported in Table\ \ref{table:photo}. We obtain a variation from 0.85 up to 1.40 with an average value of $1.16 \pm 0.20$ for the Y--H color. Even if the Y--H color of HD142527B seems to change at different epochs, the measurement errors we estimate therefore do not allow us to be confident about this color variation. This is also supported by the spectrum obtained in the night of June 13, 2016, which has a bluer color (Y--H$=0.85 \pm 0.28$) than the others, but has a later spectral type (M7V, see Table\ \ref{table:spec}) according to the fitting procedure with young field dwarf spectra. These results do not allow us to draw a firm conclusion on the spectral type of HD142527B.\ It could be both later or earlier than the M5--M6 spectral type stated here.
%Notably, the spectrum taken in the 2016--06--13 night is redder then the other spectra. This spectral and photometrical variation could be due to: i) variation of the stellar temperature; ii) variation of the contribution due to the accretion disk around the secondary and iii) variation within the circumstellar material absorption (around the primary or around the secondary, or both). \rev{Actually, because all of these variations, or part of them, could be at work in any moment making redder the real spectrum of HD 142527 B, we take into account that the latter could be earlier than the M5 -- M6 stated here}.
%For these reasons, we consider the real spectral type of HD142527B obtained here as the result of many factors that likely make it later than the real spectral type of HD142527B that could be earlier than that stated.
%For these reasons,  the spectral type obtained here as the result of many factors that likely make it later than the real spectral type of HD142527B.

\subsection{Photometry}
\label{sec:photo}
We estimated the wide-band photometry in the Y, J, and H spectral bands by adopting a distance of 156 pc and using the magnitude for the star obtained from 2MASS \citep{cutrietal20032mass}. We combined the median contrasts obtained for all the wavelength channels below 1.15 \mic for the Y band, between 1.15 and 1.35 \mic for the J band, and between 1.35 and 1.65 \mic for the H band. We derived median contrasts for the H2 and H3 bands from the July 2015 non-coronagraphic SAM image, and median contrasts for the K1 and K2 bands were derived from the May 2017 and April 2018 dataset. The uncertainty on the contrast was then calculated from the median on the error on each single channel in the wavelength ranges defined above.  These values were then transformed into the absolute magnitude for HD142527B and are listed Table~\ref{table:photo}, where the error bars are calculated taking into account the uncertainties on the distance of the system reported in Section~\ref{sec:intro}. 
From the values listed in Table~\ref{table:photo}, we obtain Y$=5.34 \pm 0.21$\ mag, J$= 5.10 \pm 0.17$\ mag, and H$=4.23 \pm 0.19$ mag as mean values for the absolute magnitudes of the secondary.

In the determination of the absolute magnitudes of HD142527B, we considered that the interstellar reddening correction for the primary and the secondary are identical, that is, assuming for the system a visible extinction of $A_V=0.6$ \citep{Verhoeff2011, lacouretal2016} and a dust reddening of $R_V=3.1$ as obtained by \cite{weingartneranddraine2001}, we obtain a value of $E(B-V)=0.194$.
From the analysis of the SED of HD142527B, \citet{lacouretal2016} stated that the emission in J or bluer bands are almost exclusively ($>90$\ \%) due to the stellar surface. The remaining 10\%, which is due to the circumstellar emission, mainly affects the J band, but the Y band is assumed to be unaffected.
%we do not consider the correction for the interstellar reddening because our measurements come from the measured contrast between the primary and its companion. 
However, the red color of HD142527B (J-H$=0.87 \pm 0.25$\ mag) is consistent with the presence of an optically thick circum-secondary disk around HD142527B itself, as has previously been identified by \citet{lacouretal2016} and \citet{2012ApJ...753L..38B}. Furthermore,  \citet{closeetal2014}, were able to detect HD142527B by means of H$_\alpha$ ($0.656\ \mu$m) observations that also allowed them to evaluate the mass accretion rate onto HD142527B ($5.9 \times 10^{-10}$\ M$_\odot$\ yr$^{-1}$), which  is about 1\% of the accretion onto the primary star \citep{closeetal2014}. Both these contributions (the disk and the accretion) are significant at longer wavelengths. From the NIR photometry of \citet{2012ApJ...753L..38B}, we can assume, as Close et al. did, a $\sim 0.3$\ mag disk H excess.   To correct the absolute magnitudes for these contributions, we used the \citet{cardellietal1989} algorithm and obtained a J-band correction of 0.1\ mag and 0.3A$_H=0.09$\ mag for the H band.
To summarize, the absolute magnitudes of HD142527B, taking into account the circumsecondary material discussed before, become Y$=5.34 \pm 0.21$\ mag, J$_{coor}= 5.20 \pm 0.17$\ mag and H$_{corr}=4.31 \pm 0.19$\ mag. 
%we must to take into consideration the red excess due to the dust emission from the circum--secondary disk identified by \citet{lacouretal2016}, \citet{2012ApJ...753L..38B} and \citet{closeetal2014}. This contribution is significant at longer wavelengths while the Y band emission is more likely emitted by the stellar surface. In order to take into account the circum--secondary disk excess, the absolute magnitude 
%reported in Table\ \ref{table:photo} should be corrected for a contribution of 0.1 mag \citep{lacouretal2016} in the J band and of 0.3 mag \citep{{2012ApJ...753L..38B}, {closeetal2014}} in the H band. 

%It is worth to point out that, contrary to \citet{lacouretal2016} and \citet{rodigasetal2014},  \citet{avenhausetal2017} did not report any flux enhancement due to nearby dust with their ZIMPOL observations, even though they reached a better contrast than the observations reported in \citet{rodigasetal2014}.
It is worth pointing out that the ZIMPOL polarimetric observation made by  \citet{avenhausetal2017} did not find any flux enhancement that would have been due to a dust disk around HD142527B, even though they reached a better contrast than the observations reported in \citet{rodigasetal2014}, where such a disk is observed. Therefore Avenhause et al. did not confirm the presence of the circum-secondary disk. However, considering a simple model of a reflective optical disk with a radius $\leq 1/3$ R$_{hill}$ around the secondary, hence at about 13 au (80 mas at 156 pc) from the primary, it is possible to estimate that the contrast between the primary and the disk is about 11 mag. This contrast value is below the $3 \sigma$ limit for the observation of HD142527 presented by \citet[panel a of their Figure 2]{avenhausetal2017}.  
%From the values listed in Table~\ref{table:photo} we obtain Y$=5.34 \pm 0.21$\ mag, J$= 5.10 \pm 0.17$\ mag and H$=4.23 \pm 0.19$ as mean values for the absolute magnitudes of the secondary that, with J and H values corrected for the contribution just discussed they become: Y$=5.34 \pm 0.21$\ mag, J$_{coor}= 5.20 \pm 0.17$\ mag and H$_{corr}=4.53 \pm 0.19$.

These magnitudes suggest some degree of variability between the different epochs, with a peak-to-valley excursion of 0.52 in the Y band and about 0.4 and 0.5 mag in the J and H band. It is not clear if this is a real effect or an artifact of our reductions. Since we worked differentially with respect to the primary, we should take into consideration the possibility that this is a variable star. Primary variability was noted by \citet{2012ApJ...753L..38B}, who reported a significant variation between the 2MASS photometry and the \citet{malfaitetal1998} photometry.
HD142527 is a Herbig Ae object that is listed as probable $\delta$ Scuti pulsator in \citet{marconiandpalla1998}. However, \citet{kurtzandmuller2001} did not find any $\delta$ Scuti pulsation, but a stellar variability with a period of about 6.0\ d. We observed the  star\footnote{At the York Creek Observatory, Georgetown, Australia} (Messina \& Millward, 2017 Priv. Comm.) from April 2017 to August 2017 for a total of 13 nights, confirming the period of 6.0\ d with a peak-to-valley amplitude of 0.13\ mag in B and 0.09\ mag in V and R. We conclude that the photometric variability amplitude of the primary is not sufficient to explain the amplitude of the variation we found in our observation of B. On the other hand, in an extensive NIR photometrical study of the RCW38 star-forming region, \citet{dorretal2013} found that most of the low-mass stars in their sample exhibit irregular light curves with typical timescales of a few days and amplitudes between 0.1 and 0.4 mag in K band. Some of them show variations and outbursts with amplitudes above 1 mag. If this is the case for HD142527B, as it may be because its very young age (1.0 - 5.0 Myr), the  variable behavior of NIR photometry of young stars can most likely explain the photometrical variation shown in Table~\ref{table:photo}.

\begin{table}
\caption{Reduced $\chi^2$ value for the fits of the HD142527B with the cool field dwarfs of the \citet{2013ApJ...772...79A} sample. The effective temperatures of the different spectral types are taken from \citet{pecautandmamajek2013}.}             
\label{table:spec}    
\centering                         
\begin{tabular}{c c c c c} 
\hline\hline                
Date & Sp &T$_{eff}$  & Reduced $\chi^2$ \\    % table heading 
\hline                        % inserts single horizontal line
2015-05-13 & M6V$^{a}$ & 2850 $\pm$ 200  & 0.443   \\
2015-07-03 & M5V$^{b}$& 3030  $\pm$ 200  & 1.013   \\
2016-03-26 &  M5V$^{b}$& 3030  $\pm$ 200 & 1.760  \\
2016-06-13 &  M7V$^{c}$ & 2650  $\pm$ 200 & 0.419 \\
2017-05-16 &  M6V$^{a}$& 2850  $\pm$ 200 & 1.035  \\
\hline  
%\\                                 %inserts single line
\multicolumn{2}{l}{$^a$ 2MASSJ04221413+1530525}\\
\multicolumn{2}{l}{$^b$ 2MASSJ06195260--2903592}\\
\multicolumn{2}{l}{$^c$2MASSJ05575096--1359503}\\
\end{tabular}
\end{table}

\begin{figure*}%[!htp]
\begin{center}
\centering
\includegraphics[width=8.0cm]{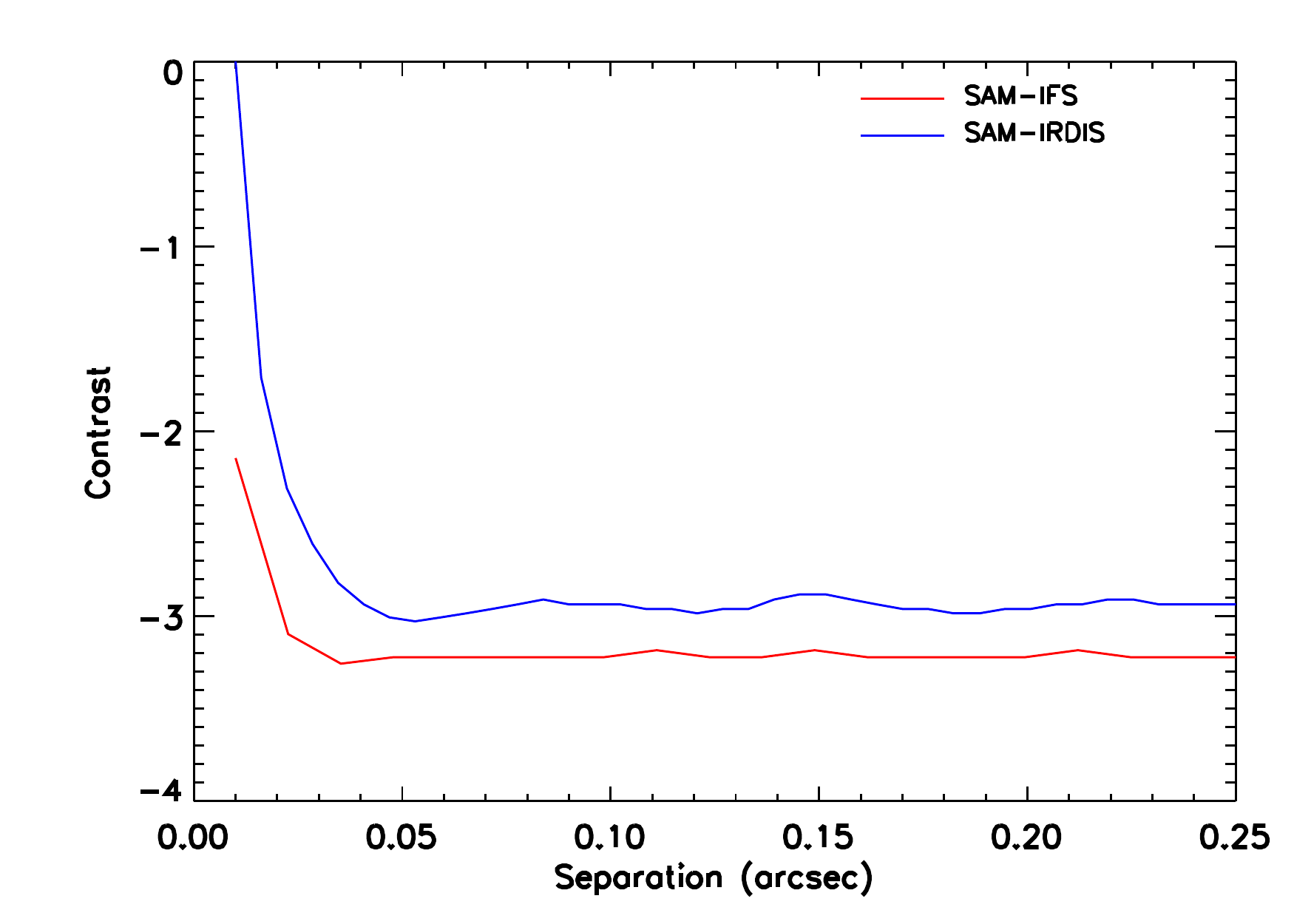}
\includegraphics[width=8.0cm]{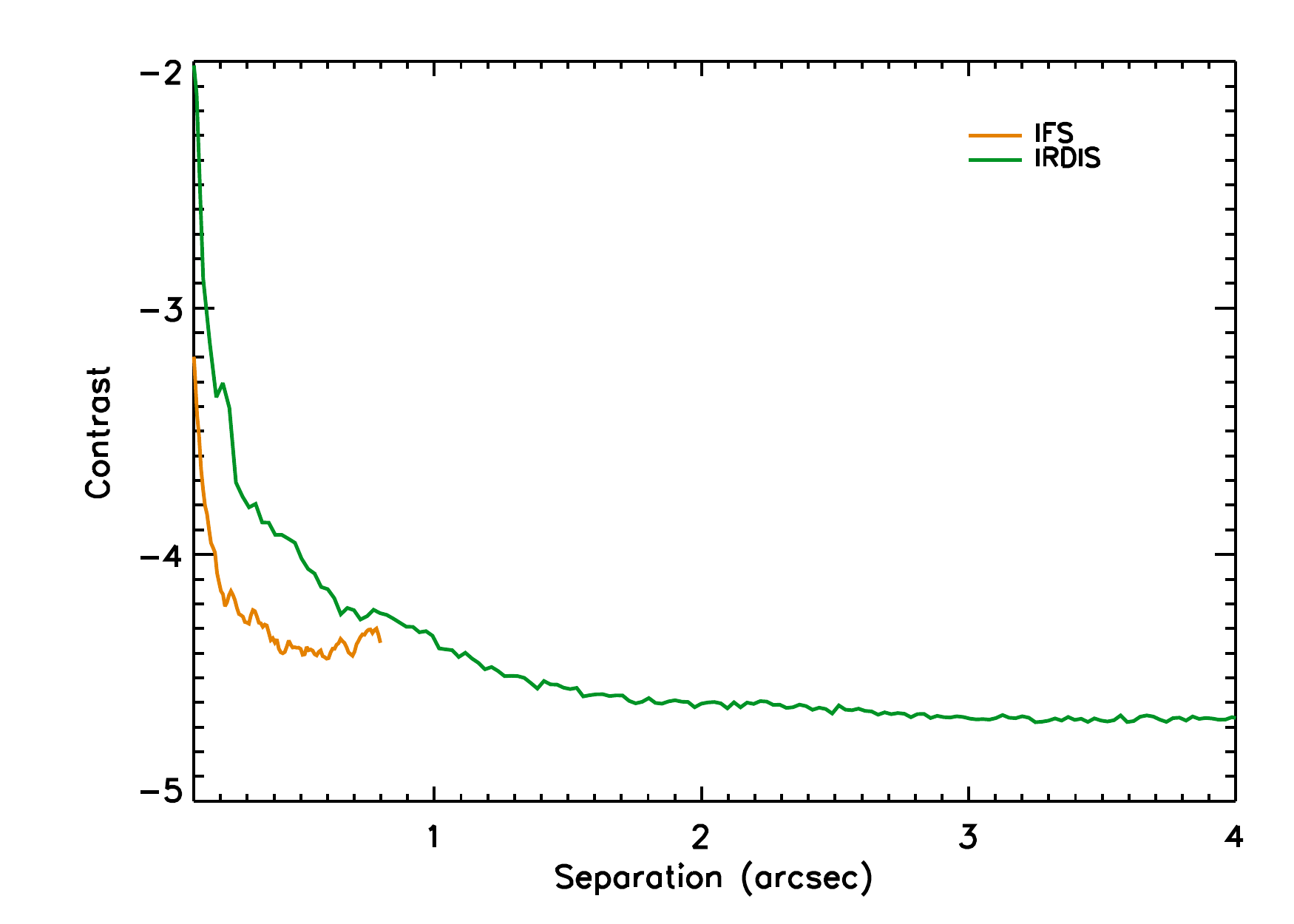}

\includegraphics[width=8.0cm]{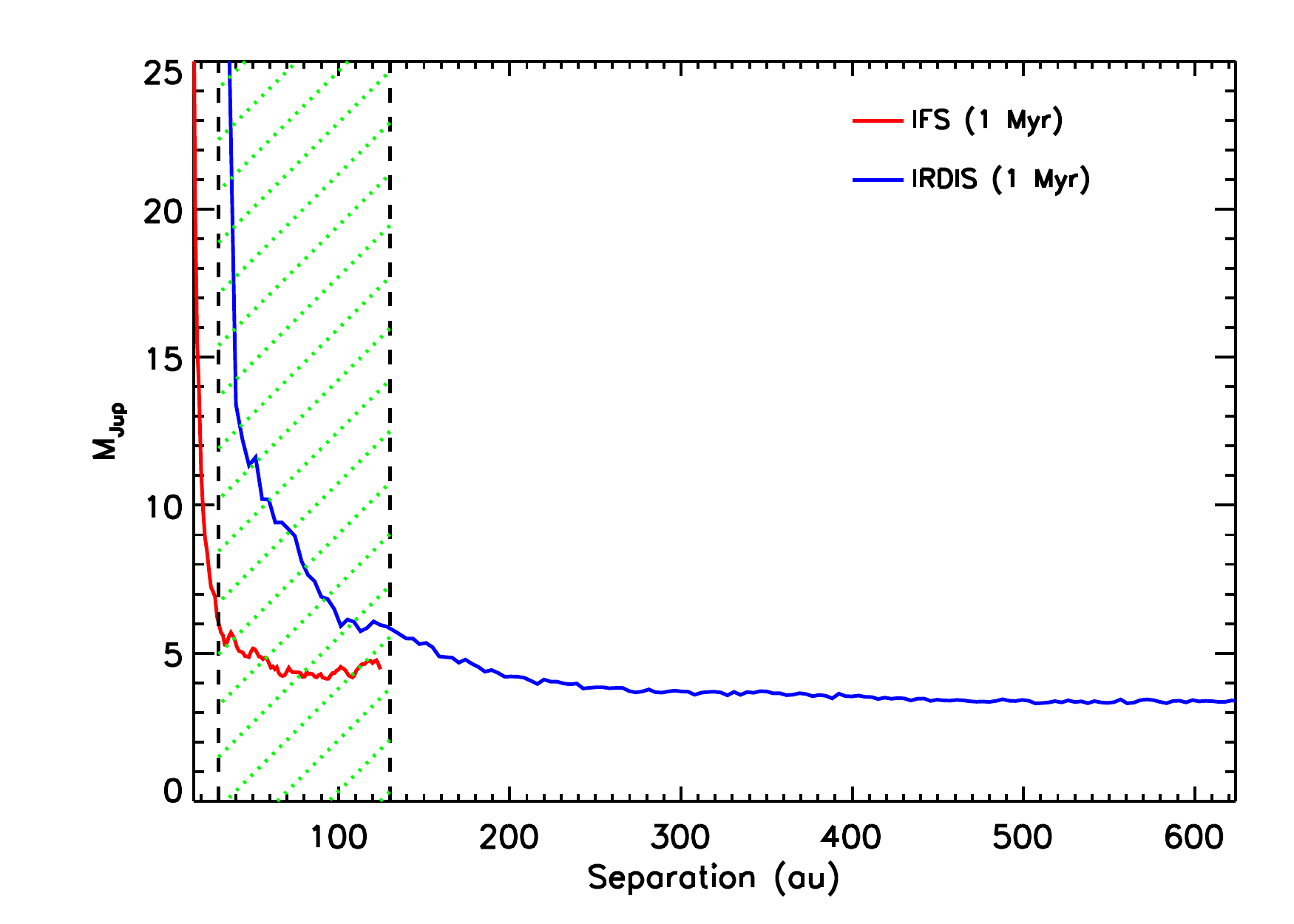}
\includegraphics[width=8.0cm]{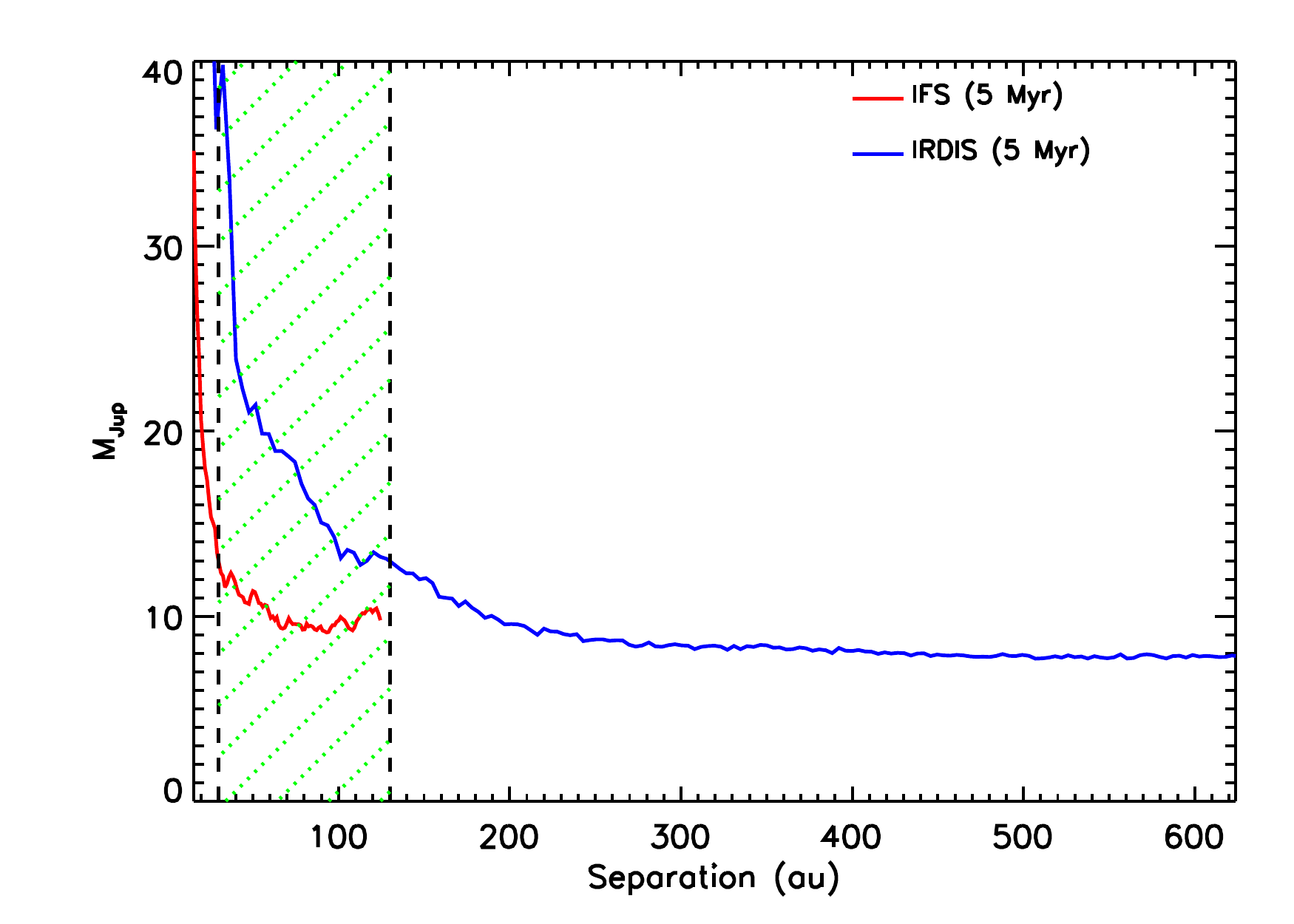}
\caption{{\bf Top left}: IFS and IRDIS contrast limits achieved within 250 mas using the SAM observing mode.  {\bf Top right}: IFS and IRDIS contrast limits achieved during the non-coronagraphic observations of HD142527. {\bf Bottom left}:  Minimum detectable mass as a function of separation in the field of view of IFS and IRDIS, adopting an age of 1 Myr and the models of BHAC2015.  {\bf Bottom right}: Minimum detectable mass as a function of separation, adopting an age of 5 Myr.  The shaded area between the two dashed vertical lines indicates the disk gap as reported by \citet{Verhoeff2011}.}
\label{f:mlimits}
\end{center} 
\end{figure*}

\subsection{Astrometry}
\label{sec:astro}
To precisely determine the position of the companion from the IFS images taken at different observing epochs, we inserted negative-simulated companions (built from a 2D Gaussian function) at different positions around the position of the true companion.  The best fit to this position was found from the position of the negative simulated companion, which minimized the standard deviation in the region around it.  The results for each epoch are listed in Table~\ref{table:astro}. 
%This procedure was executed for different images obtained with different numbers of principal components in the PCA and the error adopted on the position was calculated as the standard deviation of these measurements. 
Since the observations are non-coronagraphic, the position of the primary star could be defined with high precision because its signal is strong and unbiased. This shows that the dominant source of uncertainty remains the centering of the secondary star, which is assumed to be accurate to half of a pixel scale. The main contribution to the error on the position
angle is dominated by the uncertainty on the true north (TN) angle, calculated by observing an astrometric calibration field \citep{maireetal2016b}.

\begin{table*}
  \caption{Astrometric position derived for HD142527B for each individual epoch.}             
\label{table:astro}    
\centering                         
\begin{tabular}{c c c c c} 
\hline\hline                
Date & $\Delta$RA           & $\Delta$Dec            & Separation &  Position Angle \\    % table heading 
         &    (mas)                 &      (mas)                 &  (mas)       &  (deg)\\
\hline                        % inserts single horizontal line
2015-05-13 &  65$\pm$2         &  $-24 \pm$2        & 69$\pm$2         & 110.2$\pm$0.5 \\
2015-07-03 & 62.9$\pm$0.4    & $-18.3 \pm 0.5$  & 65.5$\pm$ 0.4  & 106.2 $\pm$ 0.4\\
2016-03-26 &  60$\pm$2         &  $-7 \pm$2          & 60$\pm$2         &  97.1$\pm$0.5 \\
2016-06-13 &  61$\pm$2         &  $-7 \pm$2          & 61$\pm$2         &  96.3$\pm$0.5 \\
2017-05-16 & $47.4 \pm 0.5$  &$10.3 \pm 0.2$    &  48.5 $\pm$ 0.5& 77.8 $\pm$ 0.2 \\
2018-04-14 &$36 \pm 1$         & $25 \pm 1  $       & $44 \pm 1$      & $55.4 \pm 0.4$\\
\hline                                   %inserts single line
\end{tabular}
\end{table*}

\subsection{Detection limits}
The best limiting contrast curves as a function of separation obtained for both IFS and IRDIS (see \cite{mesaetal2015}) for a complete description of the method we used to evaluate the limiting contrast curves) are shown in the upper panels of Figure\ \ref{f:mlimits}. Here we also show the best limiting contrast curve obtained with the SAM observing mode for IRDIS and IFS for separations smaller than 250 mas (39\ au). The lower panel shows the minimum detectable mass as a function of separation in the field of view (FoV) of IFS and IRDIS. In the figure, the inner and outer limits of the disk gap are also highlighted with vertical dashed lines. We considered the mass limits for ages of 5\ Myr, as stated for HD142527 by \citet{Fukagawa2006}, and of 1\ Myr, which resulted from comparing the photometry of HD142527B with models from \citet[][henceforth BHAC 2015]{2015A&A...577A..42B} (see Section\ \ref{sect:evmo}). The non-coronagraphic setup that was used for the observations means that the achieved contrast at sep $\geq 0.7$ arcsec is lower than the contrast achieved in coronagraphic observations. The less favorable pixel scale of the IRDIS observations and the proximity of HD142527B to the central star produce lower contrasts at small separations than in the IFS performance. For the IFS FoV, we reach minimum detectable companion masses of $7-10$\ M$_J$ (depending on the age of the system). \citet{boehleretal2017}, in their analysis of ALMA observations of HD142527, found a compact source in the continuum map as well as CO emission at about 50\ au from the central star. They interpreted this as material orbiting a low-mass companion. Except for HD142527B, we detect no companion with a mass higher than 10 M$_J$ orbiting HD142527 in the large gap between the inner and the circumbinary disks. Consequently, the third object in the system proposed by \citet{boehleretal2017}, if it exists, should be less massive than 10\ M$_J$. At larger orbital radii up to 500 au (the complete FoV of IRDIS), no planets with masses higher than $3-7$\ M$_J$ (depending on the age of the system) are detected.

\section{HD142527B mass estimate}
\label{sec:massB}
The spectroscopic results discussed in Section\ \ref{sec:Spec} can be used to estimate the mass and radius of HD142527B. The analysis of \citet{pecautandmamajek2013} on cool pre-main-sequence stars also provides a tentative determination of the mass and radius of such objects based on spectral type. Our determination of the spectral type yields a value of M\ $=0.11^{+0.04}_{-0.02}$\ M$_\odot$ and R$=0.15^{+0.07}_{-0.01}$\ R$_\odot$.

\subsection{Comparison to evolutionary models}
\label{sect:evmo}
\begin{figure}%[!htp]
\begin{center}
\centering
\includegraphics[width=8.5cm]{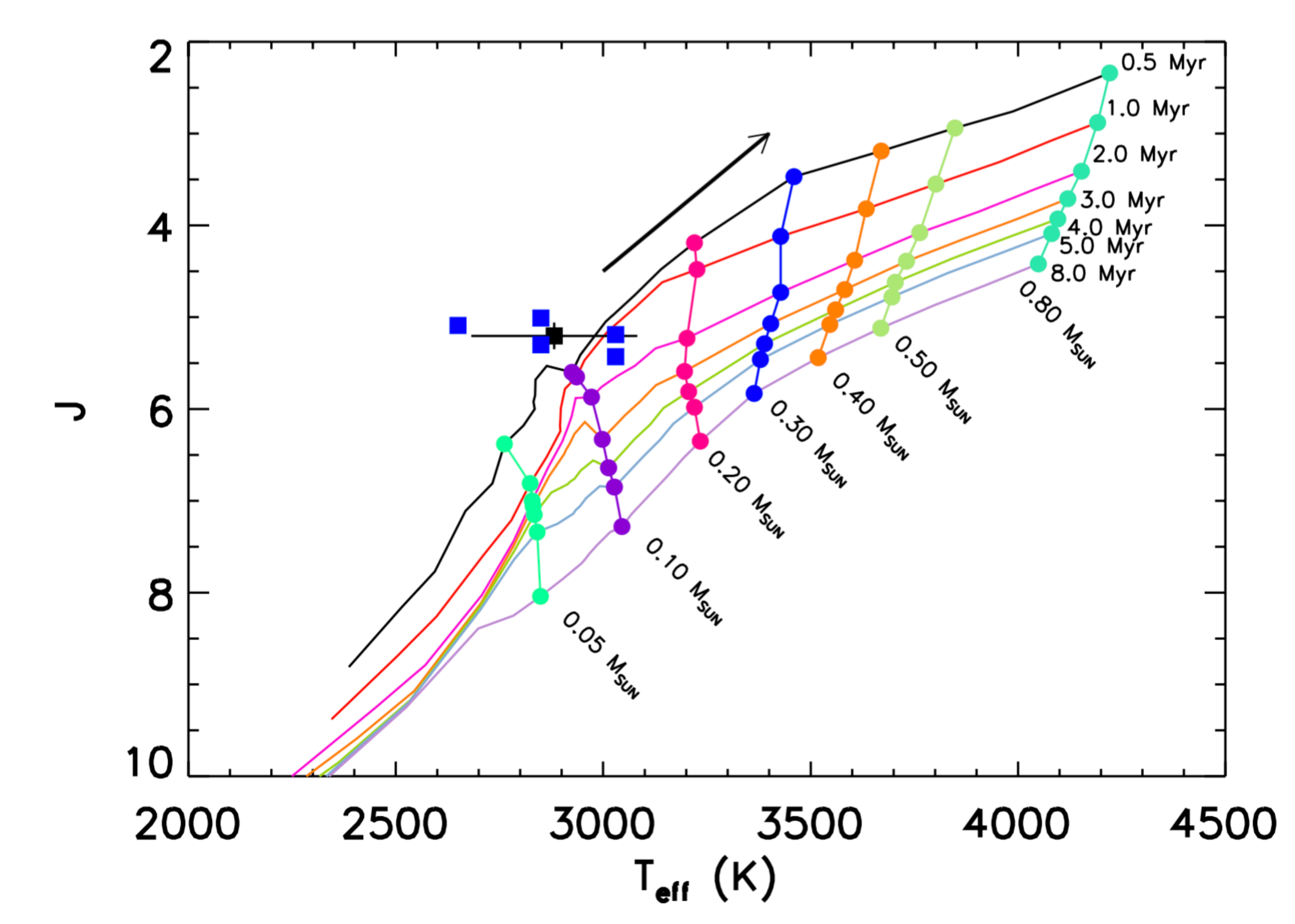}
\caption{Evolutionary tracks and isocontours of masses as function of  the temperature and absolute magnitude J as evaluated in BAHC 2015 %\citep{2015A&A...577A..42B}. 
The blue squares are the observed absolute magnitudes of HD142527B corrected for the contribution due to the circum-secondary disk, and the black square is their mean with error bars. The arrow qualitatively indicates how the isochrone system would change with respect to the observed points if the accretion process is taken into account in the models.
%the further magnitude and temperature corrections necessary if other contributions by circum -- secondary and other components of the circum--stellar environment are taken into account.
}
\label{f:teff_j}
\end{center} 
\end{figure}

The mass, radius, and age of this star can be obtained by comparisons between the measured absolute magnitudes and model absolute magnitudes from theoretical evolutionary tracks. In this case, we considered the standard evolutionary models of BHAC 2015. In Figure\ \ref{f:teff_j} we display the isochrones for models with masses ranging between 0.05\ M$_\odot$ to 0.80\ M$_\odot$ and for ages between 0.5 Myr to 8 Myr, on which we overlay the observed points (blue dots) at different epochs and their average. A suitable set of parameters for this comparison is M$=0.14 \pm 0.03$\ M$_\odot$, R$=1.2 \pm 0.5$\ R$_\odot$ , and an age of about 1 Myr. Although this mass determination agrees with our previous estimate based on spectroscopic data and with the result reported in \citet{lacouretal2016}, it is most likely not conclusive. The absolute magnitudes of HD142527B are systematically too bright compared to the models. This is mainly due to the accretion of matter onto this young star \citep{closeetal2014} that is not taken into account in the evolutionary tracks of structures, which instead only account for gravitational contraction. The accretion produces a hotter object with a larger radius. We tried to account for the contribution of the disk (see Section\ \ref{sec:photo}), but other effects can contribute to the luminosity of the secondary, such as the inner disk. It is expected that if such effects were accounted for in the isochrones, the points would shift toward larger magnitudes. If we were to take the higher temperature and brighter absolute magnitude that is due to accretion into account in the model, the isochrone system would move mainly parallel to the arrow in the Figure\ \ref{f:teff_j}, rendering this analysis inconclusive.

\subsection{Dynamical constraints}
An alternative and promising way to constrain the mass of HD142527B is to evaluate the proper motion variation of the primary star that is due to the presence of the secondary. To this purpose, we searched for significant differences in proper motion of the star as measured at different epochs (see Table\ \ref{table:Deltamu}). HD142527 is present in several catalogs, and the comparison of GAIA and Tycho2 \citep{hogetal2000} proper motions gives $\Delta \mu_\alpha=(2.14 \pm1.00)$\ mas/yr and $\Delta \mu_\delta =(1.04 \pm 1.00)$\ mas/yr. The GAIA and SPM \citep[Southern Proper Motion Catalogue]{girardetal2011} $\Delta \mu_\alpha=( 2.40\pm 3.40)$\ mas/yr and $\Delta \mu_\delta =(9.68 \pm 3.26)$\ mas/yr and GAIA and UCAC5 \citep{zachariasetal2017} $\Delta \mu_\alpha=(1.44 \pm1.9)$\ mas/yr and $\Delta \mu_\delta =(0.34 \pm 1.90)$\ mas/yr also show significant differences in proper motion.

\begin{table*}
  \caption{Values for different catalogs and epochs of the HD142527 proper motion.}             
\label{table:Deltamu}    
\centering                         
\begin{tabular}{l c c c c c c c} 
\hline\hline                
Catalog& $\mu_\alpha$    & $\delta \mu_\alpha$& $\mu_\delta$&$\delta \mu_\delta$ &p         & $\delta$p & Epoch \\    % table heading 
            &    mas/yr              &      mas/yr          &  mas/yr      &  mas/yr                 & mas &   mas     & yr       \\
\hline                        % inserts single horizontal line
%
%TGAS$^a$                & $-11.757 $&  0.077 & $-24.460$ & 0.052 &  6.40   & 0.26   & 2015.00\\
TGAS$^a$                & $-11.76 $&  0.08 & $-24.46$ & 0.05 &  6.40   & 0.26   & 2015.00\\
Hipparcos New &  $-11.19$  &  0.93    & $-24.46$  & 0.79    & 4.29    & 0.98   & 1991.25 \\
Tycho2              &  $-13.9$    &  1.00   &  $-25.5$    &  1.0     &            &           &  1991.02 \\
SPM 4.0            &  $-14.16$  &  3.42   &  $-14.78$  &  3.26   &            &           &  2006.12\\
UCAC5$^b$               & $-13.2  $  &   1.9     &  $-24.8 $  &  1.9      &            &           &  1998.50\\
\hline 
\multicolumn{8}{l}{$^a$ Tycho -- GAIA Astrometric Solution} \\                                 %inserts single line
\multicolumn{8}{l}{$^b$ Fifth US Naval Observatory Astrograph Catalog} \\                                 %inserts single line
\end{tabular}
\end{table*}

We were therefore able to use the code for orbital parametrization of astrometrically identified new systems (COPAINS, Fontanive et al. in prep.) to evaluate the characteristics of the possible companions that are compatible with the observed $\Delta \mu$.
The code uses Eq.~\ref{eq:deltamu} from \citet{makarov2005}, derived by \citet{makarov2005}, to estimate the change in a stellar proper motion that is induced by a  companion for a range of possible masses and separations, 

\begin{equation}
    \Delta\mu \leq \frac{2 \pi \Pi R_0 M_2}{\sqrt{aM_{Tot}}}.
    \label{eq:deltamu}
\end{equation}

\noindent
where, $M_2$ is the mass of the secondary, $M_{Tot}$ is the total mass of the binary, $a$ is the semi-major axis in AU, $\Pi$ is the parallax of the system in mas, and $R_0$ takes into account the orbital phase so that $R_0=\left(\frac{1+e\cos E}{1-e\cos E}\right)^{1/2}$ , where $e$ is the orbital eccentricity and $E$ is the eccentric anomaly. A fine grid of mass and separation values is explored, and the expected $\Delta\mu$ is evaluated and compared with the observed one. 
In order to properly take into account the projection effects, the code considers for each point on the mass-separation grid $10^6$ possible orbital configurations, with eccentricities drawn from a uniform distribution (a Gaussian distribution can also be used, see \citet{bonavitaetal2016} for further details and other applications of the code). 

\begin{figure}%[!htp]
\begin{center}
\centering
\includegraphics[width=8.0cm]{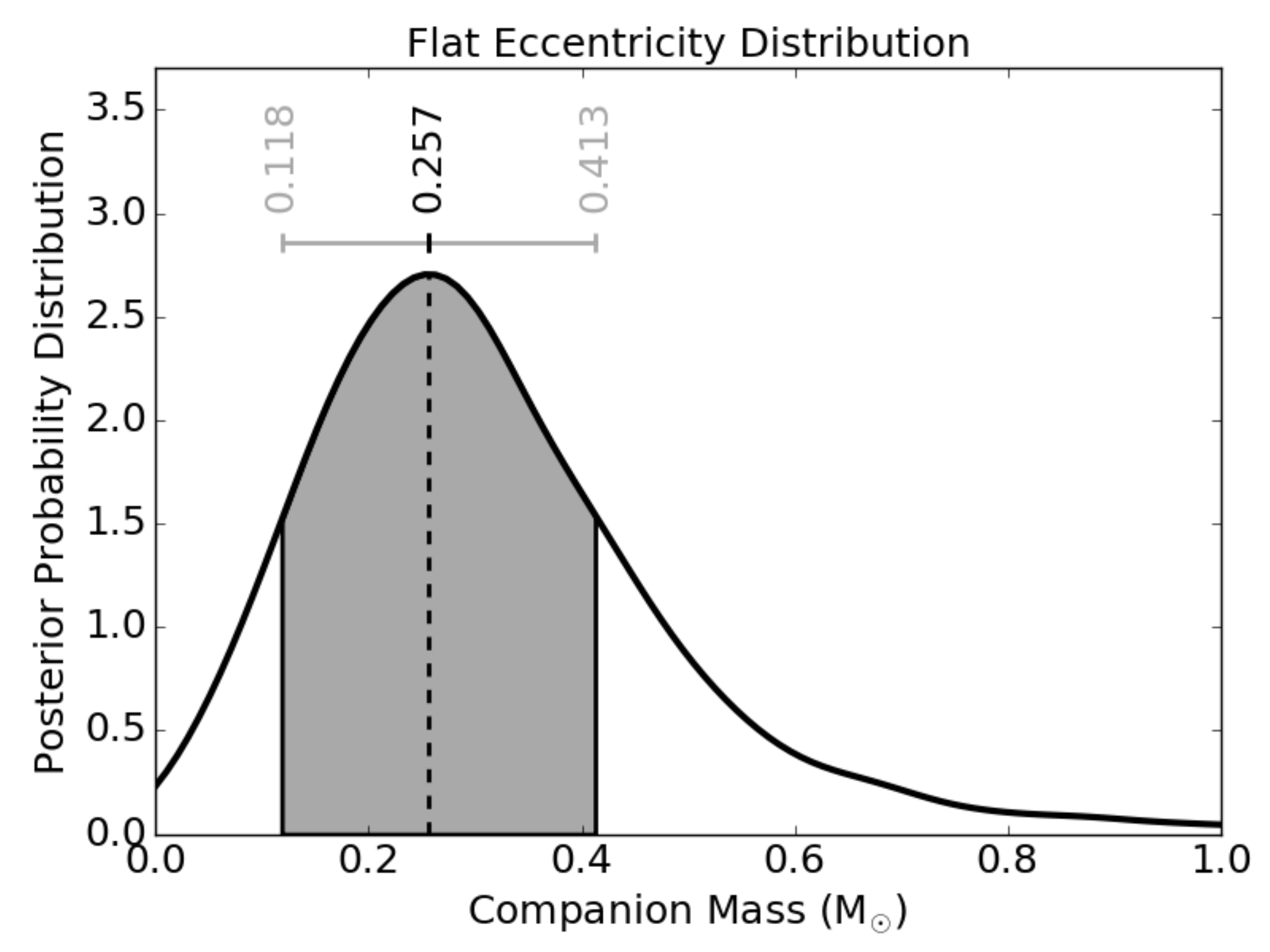}
\caption{Mass distribution, obtained using the COPAINS, of the companions that are compatible with the observed $\Delta\mu$ for HD142527, at a physical separation that is compatible with the SPHERE detection. The dashed line shows the position of the most likely value, and the shaded area highlights the region within a $1\ \sigma$ confidence level.}
\label{f:distrib}
\end{center} 
\end{figure}

If an estimate of the orbital parameters is available, as in the case of HD142527B (see Section\ \ref{sec:orbit} for the details of the orbital characterization), the code allows us to retrieve the mass distribution for companions compatible with the observed trend and the orbital characteristics. Figure\ \ref{f:distrib} shows the results of the application of this method to HD142527B.
The retrieved mass distribution peaks at $0.26^{+0.16}_{-0.14}~M_\odot$ and is therefore compatible with the value obtained in the analysis based on the spectral classification and the calibration by \citet{pecautandmamajek2013} discussed at the beginning of this section.

\section{Orbital properties of HD142527B}
\label{sec:orbit}

\begin{figure}[t]
\centering
\includegraphics[trim = 8mm 0mm 6mm 0mm,clip,width=0.30\textwidth]{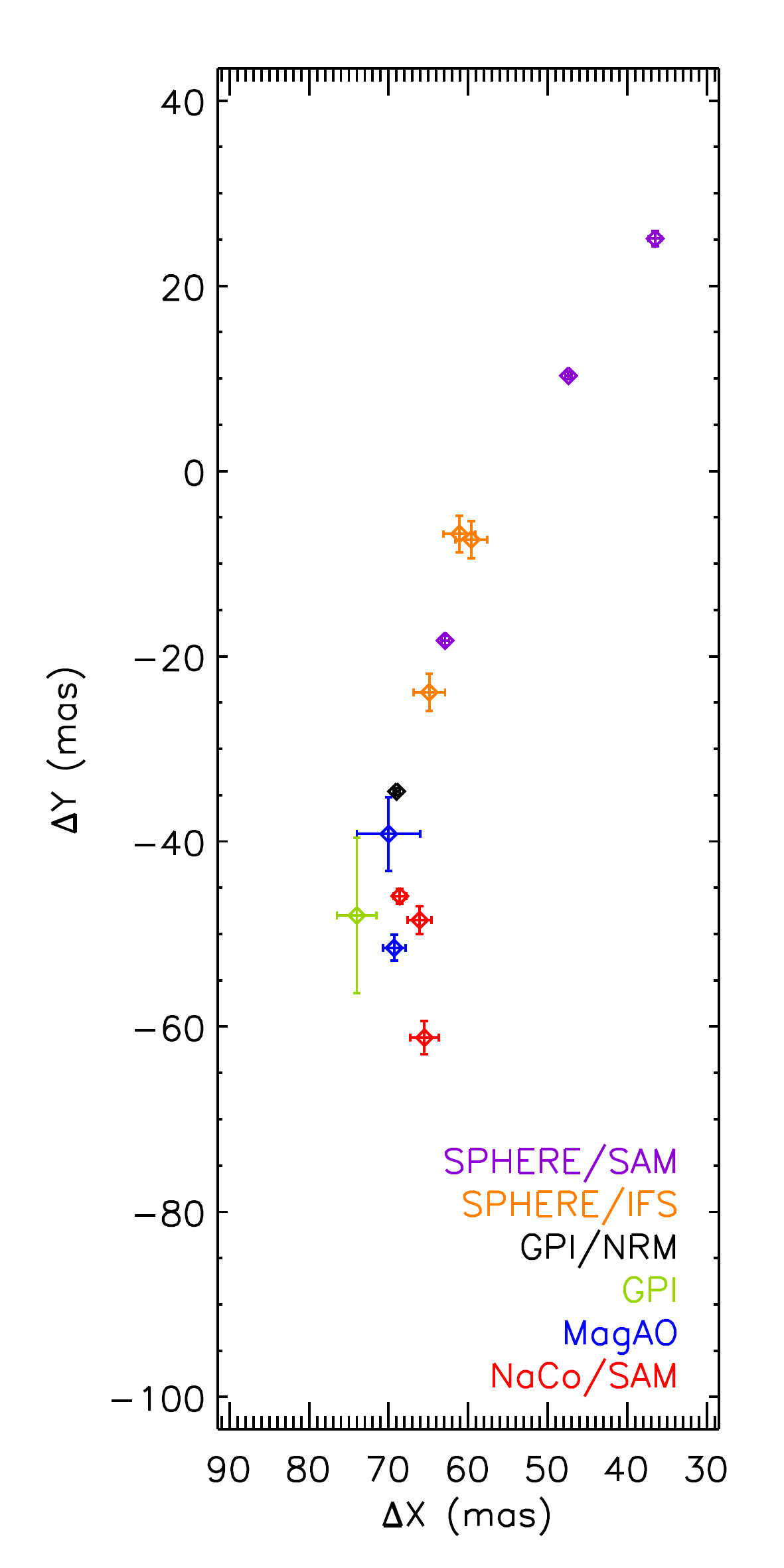}
\caption{Astrometric measurements of HD142527B from the literature (NaCo, MagAO, and GPI data) and this work (SPHERE data). The GPI measurement (green data point) is shown for comparison, but is not used in the analysis (see text).} 
\label{fig:astrometry}
\end{figure}

\begin{figure}[t]
\centering
\includegraphics[trim = 8mm 4mm 2mm 4mm,clip,width=0.50\textwidth]{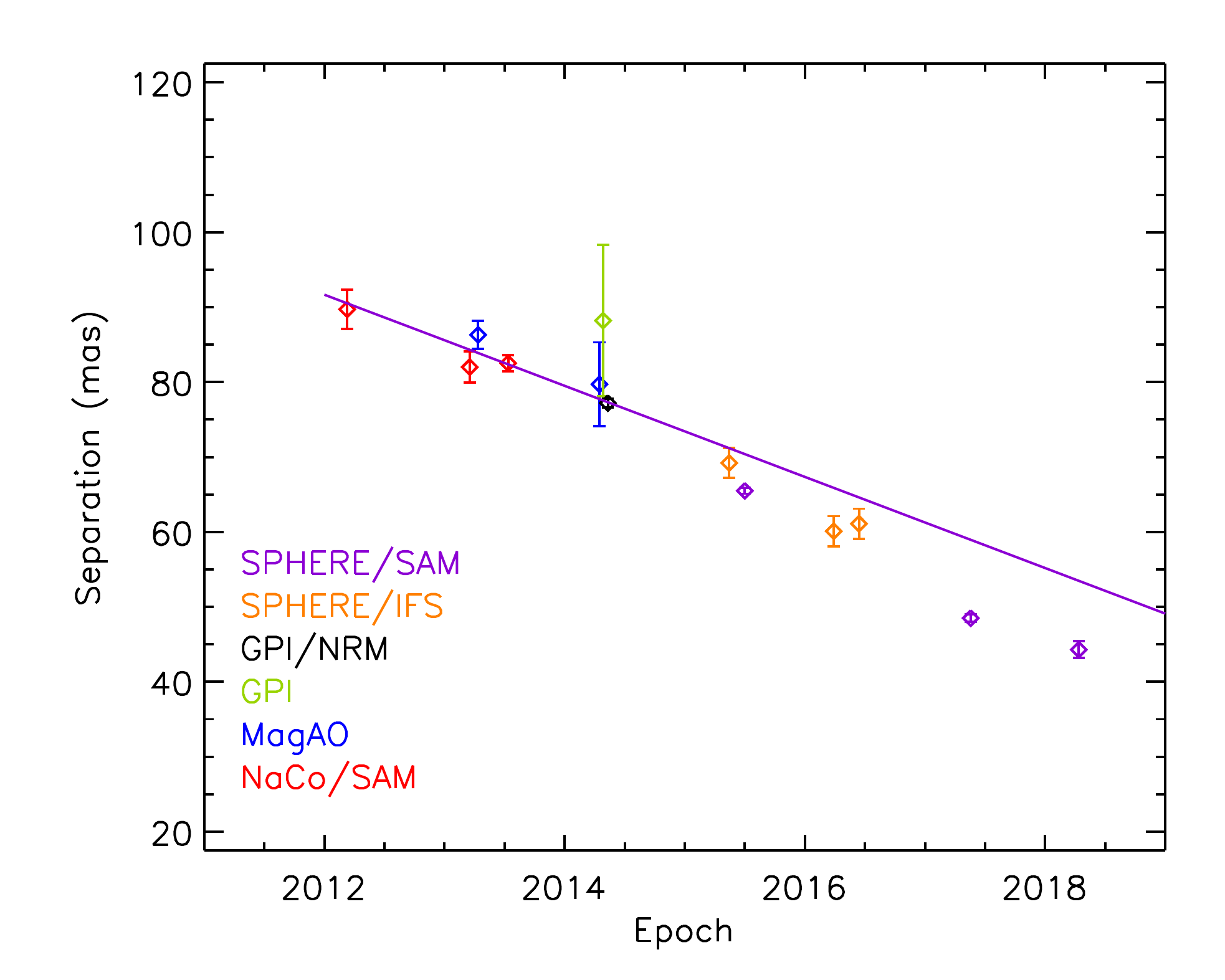}
\includegraphics[trim = 8mm 4mm 2mm 4mm,clip,width=0.50\textwidth]{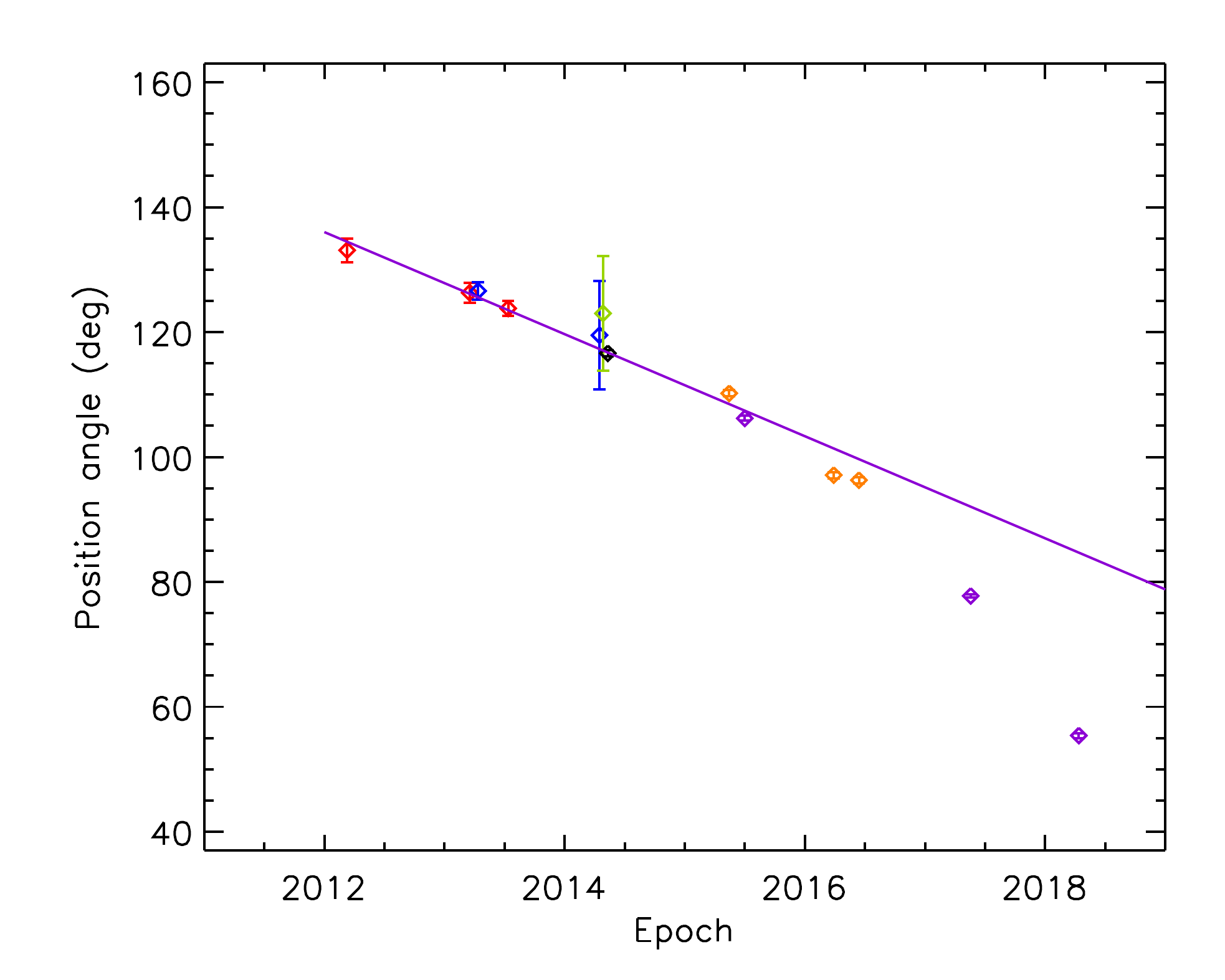}
\caption{Temporal evolution of the separation (\textit{top}) and position angle (\textit{bottom}) of HD142527B. For both panels, a linear fit to the data points obtained before 2015 is shown (purple solid line) to show the progressive deviation from linearity of the SPHERE measurements. The GPI measurements (green data points) are not considered for the fits.} 
\label{fig:seppatime}
\end{figure}

We combined the astrometric measurements of the companion reported in the literature \citep{closeetal2014, rodigasetal2014, lacouretal2016} with the new SPHERE/IFS and SPHERE/SAM measurements (Table\ \ref{table:astro}) to perform a new orbital study. The measurements are shown in Fig.~\ref{fig:astrometry}. We show the GPI polarimetric differential imaging measurement in \citet{rodigasetal2014} for comparison, but we did not use it for our analysis because of its large uncertainties and because a more accurate GPI SAM measurement close in time is available from \citet{lacouretal2016}. The SPHERE data represent an increase by almost a factor 3 in the observational baseline with respect to the previous study of \citet{lacouretal2016}. In about four years, the position angle of the companion decreased by $\sim$61$^{\circ}$ and its separation decreased by $\sim$33~mas.
%In three years, the position angle of the companion decreased by $\sim$39$^{\circ}$ and its separation decreased by $\sim$29~mas. 
The data in Fig.~\ref{fig:astrometry} indicate inflections in both separation and position angle in the orbital motion of the companion. We show the evolution of the separation and position angle of the companion as a function of time in Fig.~\ref{fig:seppatime}. The separation and position angle measured in 2018 deviate from a linear trend based on the data points listed in \citet{lacouretal2016} at a significance of  $\sim$8 and 73~$\sigma$, respectively.  In addition, the companion separation decreases. We therefore conclude that the companion is accelerating on its orbit and approaches its periapsis. 
%This is consistent with the scenario that the companion is accelerating on its orbit and getting closer to its periapsis.

\begin{figure*}[t]
\centering
\includegraphics[trim = 0mm 0mm 0mm 0mm,clip,width=0.99\textwidth]{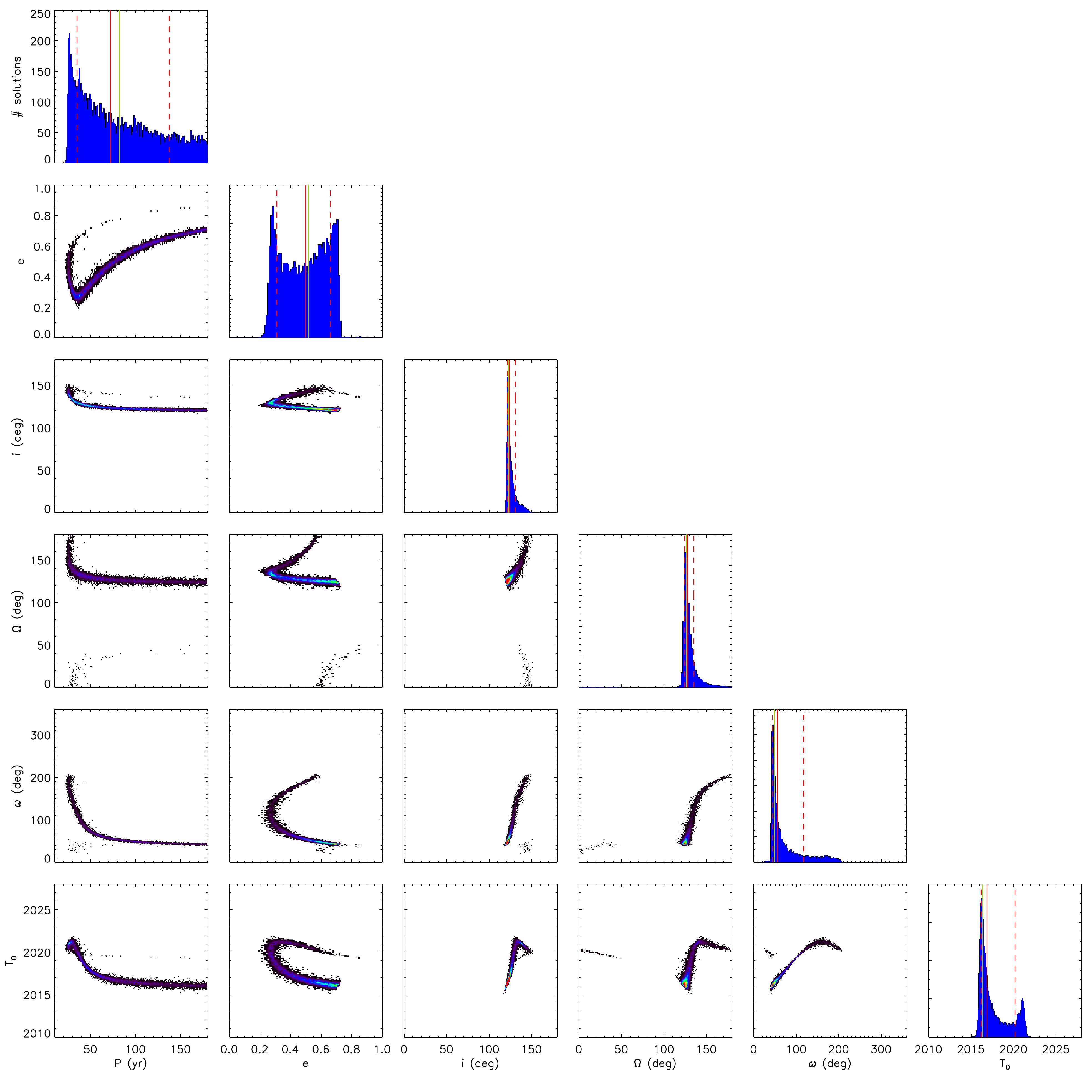}
\caption{LSMC distributions of the six Campbell orbital elements for all the fitted solutions with $\chi_{\rm{red}}^2$\,<\,2 among 2\,000\,000 random trials. The diagonal diagrams represent the 1D histogram distributions of the individual elements. The off-diagonal diagrams show the correlations between pairs of orbital elements. The linear color scale in the correlation plots accounts for the relative local density of the orbital solutions. In the diagonal histograms, the red solid line represents the 50 percentile values, the red dashed lines show the intervals at 68\%, and the green solid line indicates the best $\chi^2$ fitted solution.} 
\label{fig:cornerplot_orbital_elements}
\end{figure*}

We used a least-squares Monte Carlo (LSMC) algorithm to fit the astrometric measurements and derive distributions of the orbital parameters. The approach has previously been described in \citet{Esposito2013} and \citet{Maire2015}. We assumed a distance for the system of 156~pc \citep{GaiaCollaboration2016} and a total system mass of 2.1~$M_{\sun}$ \citep{lacouretal2016}. We drew 2\,000\,000 random realizations of the astrometric measurements assuming Gaussian distributions around the nominal values. Then, we fit the six Campbell elements simultaneously using a debugged version of the downhill simplex \texttt{AMOEBA} algorithm\footnote{The customized built-in routine provided by IDL truncates the stepping scales to floating point precision, regardless of the type of input data.} \citep{Eastman2013}. Initial guesses for the orbital elements were drawn assuming uniform distributions. We considered no priors on the orbital elements, except for the period ($P$=10--2000\,yr). Figure~\ref{fig:cornerplot_orbital_elements} shows the histogram distributions of the orbital parameters for all the derived solutions with $\chi_{\rm{red}}^2$\,<\,2.  The 68\% intervals for the parameters are orbital period $P$\,=\,35--137~yr, inclination $i$\,=\,121--130$^{\circ}$, longitude of node $\Omega$\,$\sim$\,124--135$^{\circ}$, and argument of periapsis $\omega$\,=\,44--117$^{\circ}$. For the eccentricity and time at periapsis passage, each distribution exhibits two groups of possible values: $\sim$0.2--0.45 and $\sim$0.45--0.7 for $e$, and $\sim$2015--2020 and $\sim$2020--2022 for $T_0$.
%The 68\% intervals for the parameters are: orbital period $P$\,=\,29--120~yr, inclination $i$\,=\,122--138$^{\circ}$,  longitude of the ascending node $\Omega$\,$\sim$\,122--137$^{\circ}$, and the argument of periapsis  $\omega$\,=\,46--137$^{\circ}$. For the eccentricity and time at periapsis passage, each distribution exhibits two groups of possible values: $\sim$0.2--0.45 and $\sim$0.45--0.7 for $e$, $\sim$2015--2018.5 and $\sim$2018.5--2022 for $T_0$.

\begin{figure*}[t]
\centering
\includegraphics[trim = 8mm 4mm 2mm 4mm,clip,height=0.28\textwidth]{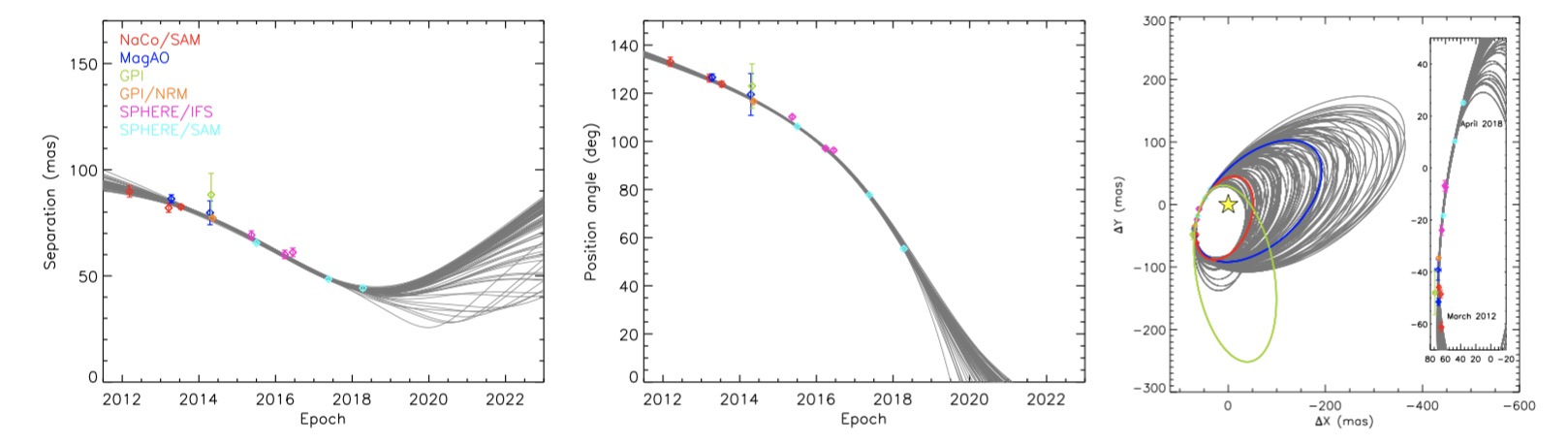}
\caption{Predicted separations (\textit{left}), position angles (\textit{middle}), and sky-plane positions (\textit{right}) for the subset of 100 randomly selected orbital solutions shown in Fig.~\ref{fig:cornerplot_orbital_elements}. In the right panel we also show with different colors three representative orbits among the fitted solutions (see text) and an inset providing a zoom around the region that is covered by the data points.} 
\label{fig:seppatimepredic}
\end{figure*}

The shape of the distributions of the orbital parameters are broadly consistent with those derived in \citet{lacouretal2016} using a Markov chain Monte Carlo algorithm for the parameters in common, except that all the solutions are consistent with a companion approaching its periapsis passage. Some parameters are better constrained by our updated analysis (inclination, longitude of node, and argument of periapsis), whereas others have larger ranges (period). The latest data points probe very close to the periapsis passage, hence we cannot firmly conclude whether the companion has passed it. Further monitoring is required to address this point. \citet{lacouretal2016} found two families of orbital solutions, where the companion approaches periapsis in one case and has recently past periapsis in the other case. They were not able to distinguish between these two solutions because available data offer only small orbital coverage and the evolution of the separation and position angle with time does not deviate from linearity. Figure~\ref{fig:seppatimepredic} shows the predicted separations and position angles for all the orbital solutions derived in our analysis. The position angle (PA) decrease will continue in the coming years with a high variation rate, while the separation will reach minimum before increasing again starting from $\sim$2019. Astrometric follow-up of the companion will be critical to precisely determine the time and separation at its periapsis passage.

The right panel of Fig.\ref{fig:seppatimepredic} represents the fitted orbits projected on the plane of the sky. The inset shows a zoom around the astrometric measurements. In the main panel we also indicate with different colors three representative orbits among the orbital solutions. The orbits similar to the orbit shown in blue compose the largest group of all three groups ($\sim$77\%  of all fitted orbits). They are characterized by a longitude of node larger than 100$^{\circ}$ and periods longer than $\sim$40~yr. The orbits represented by the orbit marked in red are the second largest group ($\sim$22\% of all fitted orbits) and have longitudes of node larger than 100$^{\circ}$ , but periods shorter than 40~yr. Finally, there is a small group of orbits shown by the orbit colored in green ($\sim$0.22\%) with periods longer than $\sim$40~yr, but longitudes of node smaller than 100$^{\circ}$. These three types of orbits also have different times at periastron passages, the first group have $T_0$ before $\sim$2019, the second group around 2019.5,  and the last group after after $\sim$2020. The plot shows that they quickly diverge after the last measurement, so that further astrometric measurements in the coming years will provide strong constraints for distinguishing between these orbits.

\begin{figure*}[t]
\centering
\includegraphics[trim = 8mm 4mm 6mm 4mm,clip,width=0.42\textwidth]{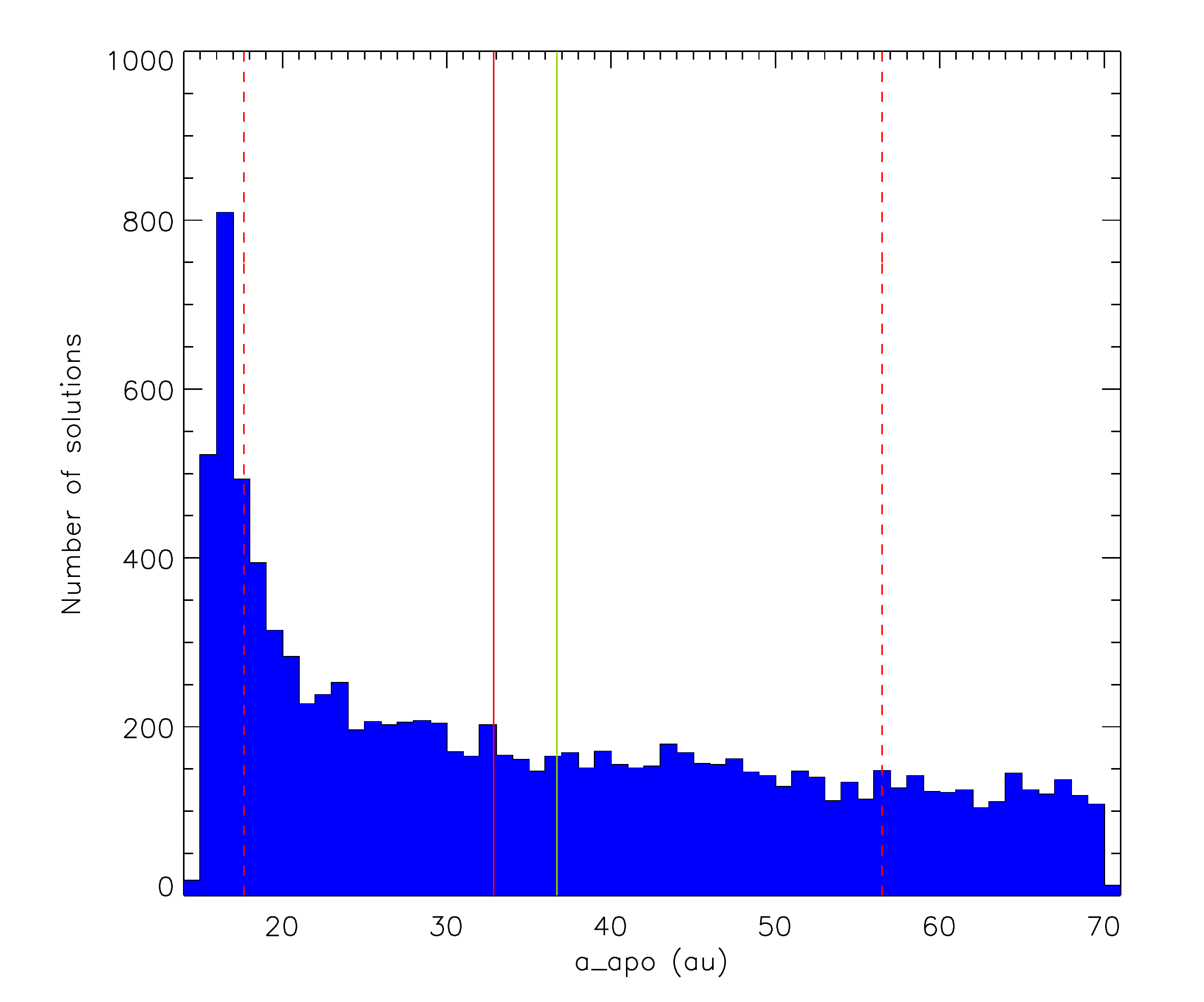}
\includegraphics[trim = 8mm 4mm 6mm 4mm,clip,width=0.42\textwidth]{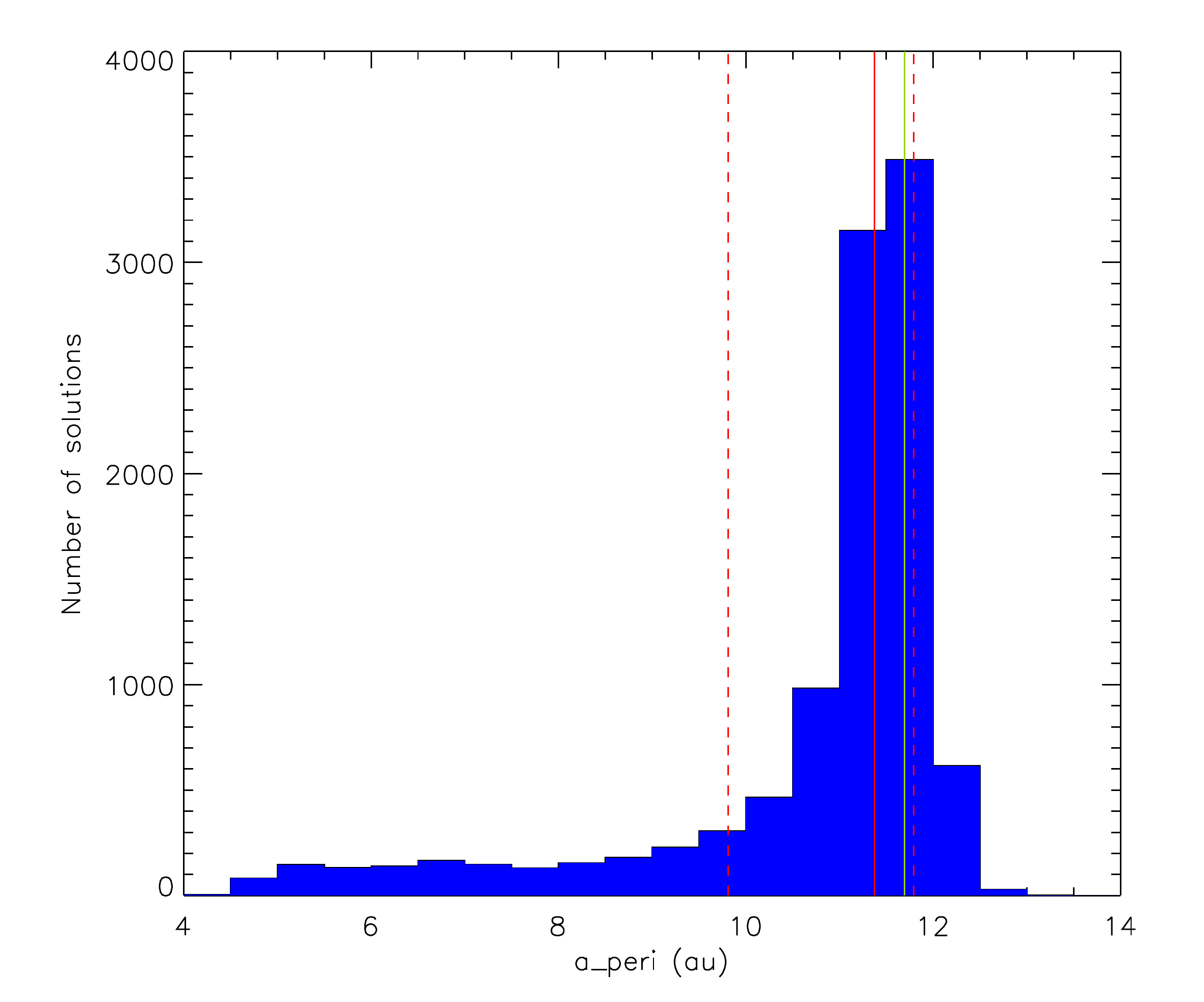}
\caption{Same as the diagonal plots in Fig.~\ref{fig:cornerplot_orbital_elements}, but for the separation at apoapsis (left) and periapsis (right).} 
\label{fig:hist_a_peri_a_apo}
\end{figure*}

The distributions of separations at apoapsis and periapsis from our LSMC analysis shown in the left panel of Fig.~\ref{fig:hist_a_peri_a_apo} indicate a range at 68\% $\sim$18--57~au. If the companion is coplanar with the outer disk, a $\sim$0.25-$M_{\odot}$ companion 
%of $\sim$16--51~au. A $\sim 0.25\,M_{\odot}$ companion 
with a 50\,au apoapsis and $e \sim 0.5-0.7$ would in principle create a region of orbital instability extending out to $\sim\,100\,$au \citep{holmanandwiegert1999}, which is the innermost possible location for the outer disk \citep[e.g.,][]{Fukagawa2006, Casassus2012, Rameau2012, rodigasetal2014, avenhausetal2017}. Nevertheless, the inclination and longitude of node of HD142527B disagree with those of the outer circumstellar disk \citep[$i$=28$^{\circ}$, $\Omega$=160$^{\circ}$,][]{Verhoeff2011, perezetal2015}, as has been reported by \citet{lacouretal2016}. This result would at first sight rule out that HD142527B is responsible for the outer disk truncation. However, recent hydrodynamical simulations have shown that for an eccentric companion with an almost polar inclination to the outer disk that approaches its periapsis passage, the interactions of companion and disk can reproduce several of the main observed disk features, such as its large cavity, and with the correct position angles, its spiral features and shadows \citep{priceetal2018}.
This orbital configuration is broadly consistent with the results from our orbital analysis. On the other hand, the orbital plane of the companion is close to the plane of the inner circumstellar disk, as previously suggested by \citet{lacouretal2016}. The inner disk has a position angle of 110$\pm$10$^{\circ}$ from CO(6-5) kinematics measured with ALMA \citep{Casassus2015}. Its mean radius is estimated to be about 10~au from near-infrared interferometric observations (Anthonioz et al., in prep.). From MIR imaging and SED modeling, \citet{Verhoeff2011} derived a maximum radial extension for the inner disk up to 30~au. However, \citet{Avenhaus2014} imaged  the inner circumstellar environment down to $\sim$0.1$''$ (15~au) in polarized scattered light with NaCo, but did not detect traces of an inner disk. More recently, \citet{avenhausetal2017} imaged the disk in visible polarized light with SPHERE down to 25~mas ($\sim$4~au) and found evidence for dust scattering close to the star, although it remains unclear if this scattering is related to the inner disk because of geometrical discrepancies with predictions from a modeling of the shadows projected onto the outer disk \citep{Marino2015}. Using our orbital analysis of the companion, we derived the distribution of its separation at periapsis (Fig.~\ref{fig:hist_a_peri_a_apo}, right panel) and find a 68\% confidence interval of  $\sim10-12$~au. This results indicates that the shape of the inner disk is strongly affected by HD142527B.

\section{Discussion and conclusion}
\label{sec:disc}
We here presented a detailed characterization of the stellar companion of HD142527. We refined the previously estimated spectral and orbital characteristics of HD142527B using IFS low-resolution spectra in both IRDIFS\_EXT observing mode and also using the SAM technique without a coronagraph. Because the observations were taken in non-coronagraphic mode, the IRDIS images taken in IRDIFS\_EXT mode were used not for photometry, but only to detect the mass limit. Except for HD142527B, we detect no objects with masses greater than 10 M$_J$ inside the gap of the disk, based on our achieved IFS contrasts. This constrains the mass of the third object hypothesized by \citet{boehleretal2017} at 50\ au from the central star to lower than 10 M$_J$. The images obtained with IRDIS exclude the presence of planets with masses greater than 7\ M$_J$ up to 500\ au. However, the non-coronagraphic technique we used in our observations means that the achieved contrast at sep $\geq 0.7$ arcsec is lower than is typically achieved in coronagraphic SPHERE images. A better mass limit outside and inside the IFS FoV will be obtained with future SPHERE coronagraphic observations.  

Our results confirm that HD142527B is an M star with a spectral type of M5-M7 and a T$_{eff}$ between 3030 and 2650 K.  The spectra are variable in flux because of several possible factors, such as 1) the variation in stellar temperature, or 2) the contribution from the accretion disk around the secondary, and/or 3) the variation in circumstellar material absorption (around the primary or secondary, or around both) that is due to interaction with the environment and to the young age of the system. 
In comparing the absolute magnitude of the B component to the models, we find (like \citet{lacouretal2016}) that the age of the secondary is very young ($\sim 1$\ Myr), younger than the age of the primary (5\ Myr). In order to match the same age for the companion as for the A component and to make this agree with the evolutionary tracks in Figure\ \ref{f:teff_j}, we would have to add about 2.0 mag to the J magnitude, pushing the system to a distance  of about 62 pc, which is not compatible with the distance measured by \citet{GaiaCollaboration2016}, as reported in Section\ \ref{sec:hd142527}.
%Similarly to  \citet{lacouretal2016}, by comparing its asbolute magnitude to models, we also find that the age of the companion is very young ($\sim 1$\ Myr), younger than the age of the primary (5\ Myr). If, in the model comparison, we adopt 5\ Myr for the age of HD142527B we find an apparent magnitude that is fainter than expected and would imply a distance to this system of $>156$\ pc. 
As described in Section\ \ref{sect:evmo}, this discrepancy arises because accretion is not taken into account in the young-object models. This accretion will cause the object to appear younger and redder than a non-accreting object. Because of these caveats, we consider the comparison to the theoretical models of young objects inconclusive. This means that the spectral type of the secondary star may also be earlier than M6. By studying the proper motion of the primary star, we dynamically constrained the mass of  the secondary to be M$_{HD142527B}=0.26^{+0.16}_{-0.14}$ M$_\odot$. This value is in agreement with our spectroscopic estimate (see Section\ \ref{sec:massB}) and with the estimate of \citet{lacouretal2016}.  Following \citet{pecautandmamajek2013} the lower and upper dynamical mass values correspond to a spectral type between M2.5 and M5.5. To our knowledge, this is the first dynamical mass determination for HD142527B that was made by exploiting the difference between nearly instantaneous and long-term motion. Significant improvements are expected after the full \textit{GAIA} dataset is available.

By combining the new SPHERE/IFS and SPHERE/SAM astrometric measurements with those reported in the literature, we constrained the orbital properties of HD142527B and obtained a period of P$=35-137$\ yr, an inclination of $i=121 - 130^\circ$\ degrees, a value of $\Omega=124-135^\circ$\ degrees for the longitude of node, and an 68\%  confidence interval of $\sim 18 - 57$\ au for the separation at periapsis. Eccentricity and time at periapsis passage exhibit two groups of values: $\sim$0.2--0.45 and $\sim$0.45--0.7 for $e$, and $\sim$2015--2020 and $\sim$2020--2022 for $T_0$. The orbit of the secondary is more inclined  than the outer circumstellar disk and seems to rule out that HD142527B is responsible for the truncation of the outer disk. On the other hand, these results are also consistent with the scenario described by \citet{priceetal2018}, in which a companion close to its periapsis in an eccentric and highly inclined orbit with respect to the outer disk could be responsible for the large cavity and other observed features and shadows. Our distribution of the orbital parameters are in good agreement with those of  \citet{lacouretal2016}. Our solutions are consistent with a companion that is approaching its periapsis passage, but the errors on our derived eccentricity and period are too large (but further monitoring is required) to conclusively determine whether the companion has passed periapsis. The next two or three years of observations will be crucial to clarify this point.

%\section{Conclusions}
%\label{sec:conc}

%Astrometric monitoring of the companion with \textit{Gaia} in combination with \textit{Hipparcos} data will be relevant given its large orbital inclination on sky.

\begin{acknowledgements}
SPHERE is an instrument designed and built by a consortium consisting of IPAG (Grenoble, France), MPIA (Heidelberg, Germany), LAM (Marseille, France), LESIA (Paris, France), Laboratoire Lagrange (Nice, France), INAF--Osservatorio di Padova (Italy), Observatoire de Gen\`eve (Switzerland), ETH Zurich (Switzerland), NOVA (Netherlands), ONERA (France) and ASTRON (Netherlands) in collaboration with ESO. SPHERE was funded by ESO, with additional contributions from CNRS (France), MPIA (Germany), INAF (Italy), FINES (Switzerland) and NOVA (Netherlands). SPHERE also received funding from the European Commission Sixth and Seventh Framework Programmes as part of the Optical Infrared Coordination Network for Astronomy (OPTICON) under grant number Rll3--Ct--2004--001566 for FP6 (2004--2008), grant number 226604 for FP7 (2009--2012) and grant number 312430 for FP7 (2013--2016).  This work has made use of the SPHERE Data Centre, jointly operated by Osug/Ipag (Grenoble), Pytheas/Lam/Cesam (Marseille), OCA/Lagrange (Nice) and Observatoire de Paris/Lesia (Paris) and supported by a grant from Labex OSUG@2020 (Investissements d'avenir ANR10 LABX56). This work has been in particular carried out in the frame of the National Centre for Competence in Research 'PlanetS' supported by the Swiss National Science Foundation (SNSF). D.M. acknowledges support from the ESO-Government of Chile Joint Comittee program ``Direct imaging and characterization of exoplanets''.  A.Z. acknowledges support from the CONICYT + PAI/ Convocatoria nacional subvenci\'on a la instalaci\'on en la academia, convocatoria 2017 + Folio PAI77170087. J.~O. acknowledges support from the ICM (Iniciativa Cient\'ifica Milenio) via the Nucleo Milenio de Formaci\'on planetaria grant, from the Universidad de Valpara\'iso and from Fondecyt (grant 1180395).\end{acknowledgements}

%\begin{thebibliography}{}

\bibliographystyle{aa} % style aa.bst   (è il format di A&A)
\bibliography{hd142biblio} % your references biblio.bib

\end{document}